%% file: review_hawkes.tex
\documentclass[11pt]{article}

\usepackage[parfill]{parskip}

\usepackage[toc,page]{appendix}

\usepackage{amsthm}
\usepackage{amssymb}
\usepackage{amsmath}
\usepackage{dsfont}
\usepackage{longtable}
\usepackage[a4paper]{geometry}
\usepackage[margin=1cm,font=small,labelfont=bf]{caption}

\usepackage{graphicx}

\usepackage{xparse}
\usepackage{authblk}

\newlength{\extrawidth}%
\newenvironment{wtable}[1][tbp]{%
  \begin{table}[#1]
    \hspace{-\extrawidth}%
    \addtolength{\textwidth}{2\extrawidth}%
    \begin{minipage}{\textwidth}%
}{\end{minipage}\end{table}}

\usepackage{floatpag}

\input{notation.tex}

\newtheorem{defin}{Definition}
\newtheorem{prop}{Proposition}
\newtheorem{theor}{Theorem}
\newtheorem{notat}{Notation}
\newtheorem{example}{Example}

\title{Hawkes processes in finance}
\author[1]{Emmanuel Bacry}
\author[1]{Iacopo Mastromatteo}
\author[1,2]{Jean-Fran\c{c}ois Muzy}
\affil[1]{\small Centre de Math\'ematiques Appliqu\'ees, CNRS, \'Ecole Polytechnique, \authorcr UMR 7641, 91128 Palaiseau, France}
\affil[2]{\small Laboratoire Sciences Pour l'Environnement, CNRS, Universit\'e de Corse, \authorcr UMR 6134, 20250 Cort\'e, France}
\date{}  % Activate to display a given date or no date

\begin{document}

\maketitle

\abstract{
	In this paper we propose an overview of the recent academic literature devoted
	to the applications of Hawkes processes in finance. Hawkes processes
	constitute a particular class of multivariate point processes that 
	has become very popular in empirical high frequency finance this last decade.
	After a reminder of the main definitions and properties that characterize
	Hawkes processes, we review their main empirical applications to 
	address many different problems in high frequency finance. Because of their
	great flexibility and versatility, we show that they have been successfully
	involved in issues as diverse as estimating the volatility at the level of transaction data,
	estimating the market stability, accounting for systemic risk contagion, 
	devising optimal execution strategies or capturing the dynamics of the full order book.
}

\section{Introduction}
\label{sec:intro}
The availability of high frequency financial data during the last decades allowed
the empirical finance to devise and calibrate models of market microstructure,
aiming at accounting for the intraday market dynamics in its finest details.
Until recently, there were only few continuous time models for the high frequency 
price variations. Most approaches relied on discrete time models that consist either
in aggregating the dynamics on intervals of a regular time grid or in considering the succession
of discrete time events like trades (these models are generally referred to as  ``trading time''
or ``business time'' models).
Hasbrouck \cite{Hasbrouck:1991}, Engle and Russel \cite{Engle:1998} in the nineties, were the first to advocate that the modeling
of financial data at the transaction level could be advantageously done within the framework
of continuous time ``point processes''.
Since then, point processes applications to finance is an ongoing, 
very active
topic in the econometric literature. We refer the reader to the recent review of 
Bowens and Hautsch~\cite{Bauwens:2009aa} on this subject.

The first type of point processes proposed in the context of market microstructure
is the ACD model introduced by Engle and Russel \cite{Engle:1998}. 
This model and its variants remains,
by far, the most used model in high frequency econometrics \cite{Bauwens:2009aa}.  
In this class of models, the process is defined by the means of its ``hazard function''
that specifies the conditional law of inter-event (or duration) intervals.
However, point processes (or counting processes) can alternatively be represented by 
their ``intensity function'' that represents the conditional probability density of the occurrence 
of an event in the immediate future (see e.g.~\cite{daley1988introduction} for a comprehensive textbook 
on point process mathematical properties).

In a pioneering work, Bowsher~\cite{Bowsher:2007aa}, recognized 
the flexibility and the advantages of using
the class of multivariate counting processes 
that can be specified by a conditional intensity vector. 
More specifically, he introduced a bivariate Hawkes processes
in order to model the joint dynamics of trades and mid-price changes of the NYSE.
Hawkes processes is a class of multivariate point processes that were introduced in the seventies
by A.G Hawkes~\cite{Hawkes:1971lc,Hawkes:1971nq} notably to model the occurrence seismic events.
They involve an intensity vector that is a simple linear function of past events 
(see e.g. \cite{Bauwens:2006,Bauwens:2009aa} for other examples of dynamical intensity point processes). 
Hawkes models are becoming more and more popular in the domain of high frequency finance.
This popularity can be explained above all by their great simplicity and flexibility, 
as anticipated by Bowsher~\cite{Bowsher:2007aa}. These models can easily account for the interaction
of various types of events, for the influence of some intensive factors (through marks) or for
the existence of non-stationarities. They are amenable to statistical inference and closed-form formulae 
can be obtained in some particular situations. Moreover since their parameters have a straightforward interpretation (notably
through the cluster representation), they lead to a quite simple interpretation
of many aspects of the complex dynamics of modern electronic markets.

In this paper, we propose a survey of recent academic studies using Hawkes processes  in the context of finance. As we already explained, there are many such studies. In order to present them in relation one to each other,
we had to group them by ``themes''. Of course, the boundaries of these themes are unclear (e.g., some of the studies belong to several themes), so the choices we have made are  unavoidably somewhat arbitrary. 
However we believe it helps capturing the ``picture'' of Hawkes models in finance.

This survey is organized as follows. Sec.~\ref{sec:process} is devoted to the theory of Hawkes processes. It introduces the main definitions and the general properties that will be used all along the paper. The following Sections focus on the applications of Hawkes processes to finance. Sec.~\ref{sec:univariate} starts with the main univariate models that can be found in the literature. That includes market activity or risk models (e.g., 1-dimensional market order flow models, extreme return models). Price models (mid-price or best limit price) are presented in Sec.~\ref{sec:price} whereas Sec.~\ref{sec:impact} is devoted to impact models. In this Section, we do not only discuss the influence of market order flows on price moves but also the problems related to optimal execution. Models that involve more order flows are presented in Sec.~\ref{sec:ob_models}. So-called level-I models (i.e., dealing ``only'' with the dynamics of the best limits) as well full order book models are discussed. Finally, various studies that did not clearly fit in any of the previous Sections are presented in Sec.~\ref{sec:other_models} (e.g. systemic risk models, high-dimensional models or news models). More materials can be found in Appendices. In Appendix \ref{sec:table}, 
all the academic works that are discussed throughout our paper and which involves numerical experiments on financial data is listed in a single table. This table summarizes some essential characteristics of the models and data used in each work.
Finally, two Appendices sum up the main results about simulation (App.~\ref{sec:simul}) and estimation (App.~\ref{sec:estimation}) of Hawkes processes.

\section{The Hawkes process}
\label{sec:process}
As mentioned in the introduction, Hawkes processes are a class
of multivariate point processes introduced by Hawkes in the early seventies
\cite{Hawkes:1971lc,Hawkes:1971nq} that are characterized by a stochastic
intensity vector. If $\cnt[t]$ is a vector of counting processes\footnote{A counting process is a stochastic process  $\{N_t\}_{t\ge 0}$, with values that are positive, integer, and increasing. By convention $N_0 = 0$.}
at time $t$, then its intensity vector $\lam[t]$ is defined heuristically 
as (see e.g.~\cite{daley1988introduction} for a rigorous definition):
\begin{equation}
  \label{eq:intensity}
  \lam[t] = \lim_{\Delta \rightarrow 0} \Delta^{-1}  \avc{ \cnt[t+\Delta] - \cnt[t]}{\filt{t} }
\end{equation}
where the filtration $\filt{t}$ stands for the information available up to (but not including) time $t$.
In the case of Hawkes processes, $\lam[t]$ is simply a linear function of past jumps
of the process as specified thereafter.

\subsection{Definition}
\label{ssec:defin}

We consider a $\nn$-variate counting-process $\cnt[t]=\{ \cnt[i][t]\}_{i=1}^{\nn}$, whose associated intensity vector is denoted as $\lam[t]=\{ \lam[i][t]\}_{i=1}^{\nn}$.
\begin{defin}[Hawkes process]
    A Hawkes process is a counting-process $\cnt[t]$ such that the intensity vector can be written as
    \begin{equation}
        \label{eq:hawkes}
        \lam[i][t] = \exo[i] + \sum_{i=1}^{\nn} \int \dd\cnt[j][t'] \knl[ij][t-t'] \, ,
    \end{equation}
    where the quantity $\exo=\{ \exo[i] \}_{i=1}^{\nn}$ is a vector of exogenous intensities, and $\knl[t]=\{ \knl[ij][t]\}_{i,j=1}^{\nn}$ is a matrix-valued kernel such that:
    \begin{itemize}
        \item It is component-wise positive, i.e., $\knl[ij][t] \geq 0$ for each $1\leq i,j \leq \nn$;
        \item It is component-wise causal (if $t<0$, $\knl[ij][t] = 0$ for each $1\leq i,j \leq \nn$);
        \item Each component $\knl[ij][t] $ belongs to the space of $\lone$-integrable functions\footnote{Strictly speaking, this definition only calls for $\knl[ij][t] \in L^{1}_{loc}$ but this is of no interest for this paper.}
    \end{itemize}
\end{defin}
\begin{notat}[Convolution]
    We can adopt a more compact notation so to rewrite Eq.~(\ref{eq:hawkes}) as
    \begin{equation}
        \label{eq:hawkes_mat}
        \lam[t] = \mu + \knl * \dd\cnt[t] \,,
    \end{equation}
    by defining the $*$ operation, corresponding to a matrix multiplication in which ordinary products are replaced by convolutions.
\end{notat}
\begin{notat}[Event times]
    By introducing the couples $\{ (t_m,k_m) \}_{m=1}^M$ , where $t_m$ denotes the time of event number $m$ and $k_m \in [1,...,\nn]$ indicates its component, Eq.~(\ref{eq:hawkes}) can also be rewritten as
    \begin{equation}
        \label{eq:hawes_expl}
        \lam[i][t] = \exo[i] + \sum_{m=1}^M \knl[i,k_m][t-t_m] \, ,
    \end{equation}
\end{notat}

\begin{figure}[htb]
    \begin{center}
        \includegraphics{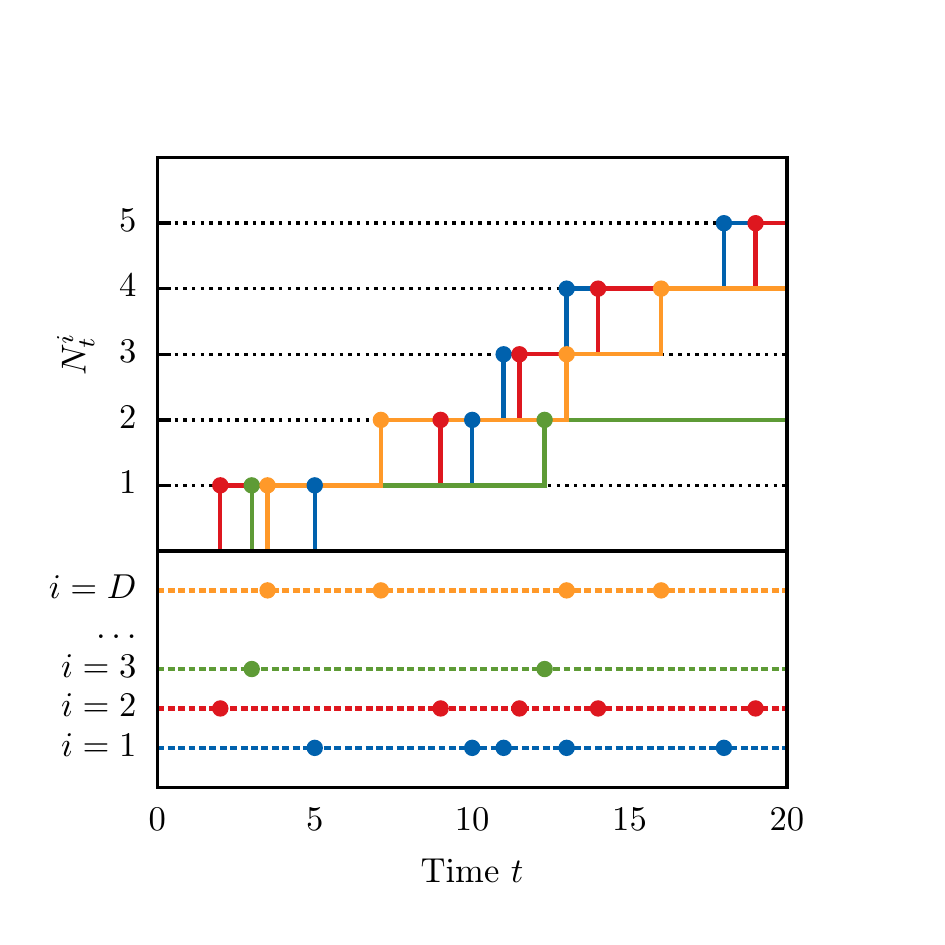}
    \end{center}
    \caption{A realization of a multivariate Hawkes process. The dots represent individual events, while the different rows refer to different $i$ coordinates.}
    \label{fig:hawkes}
\end{figure}

Fig.~\ref{fig:hawkes} shows as an example a specific realization of a multivariate Hawkes process.\\
Even though the process defined by Eq.~(\ref{eq:hawkes}) is well-defined by any choice of kernel $\knl[t]$ satisfying the three conditions stated above, the stationary case characterized below is of particular relevance in most of the applications of Hawkes processes to finance (see Sec.~\ref{ssec:extensions} for some notable exceptions).
\begin{prop}[Stationarity]
\label{prop:H}
    The process $\cnt_t$ has asymptotically stationary increments and $\lam[t]$ is asymptotically stationary if the kernel satisfies the assumption, also referred to as the {\bf stability condition}:\\
    {\bf (H)} The matrix $|| \knl ||=\{ ||\knlmat[ij]||  \}_{i,j=1}^D$ has spectral radius smaller than 1. \\
\end{prop}
\begin{notat}[Spectral radius]
    Here and in the following parts of the discussion, given a a scalar function $f(t)$ we denote with the symbol $||f||$ its $L^1$-norm, defined as $\int \dd t \, |f(t)|$. For a matrix $F=\{ f^{ij}  \}$, the notation $||F||$ will denote its spectral radius. Finally, for a matrix of functions $F(t)=\{ f^{ij}(t)  \}_{i,j=1}^D$, we will write $||F||=\{ ||f^{ij}||  \}_{i,j=1}^D$ 
\end{notat}
From now on, we will always consider (unless specified)
that assumption {\bf (H)} holds and that the
Hawkes processes are in the asymptotically stationary regime. In particular, the averages taken in the stationary state will be denoted with $\av{\dots}$, while the variances will be written as $\var{\dots}$. The consequences of the stationarity assumption {\bf (H)} will be fully explored in the next subsection. Here, we will first provide a simple implementation of a Hawkes process, whose prototypical version is the one in which the kernel functions $\knl[ij][t]$ are exponential functions.

\begin{example}[Exponential kernel]
\label{example1}
    Consider a bivariate Hawkes process in which the kernel matrix has the form\footnote{The upperscript $(s)$ (resp.\ $(c)$) stands for the word $self$ (resp.\ $cross$), since it describes the self- (resp.\ cross-) excitation of the two components.}
    \begin{equation}
        \label{eq:kernel_bivariate}
        \knl[t] = \left(\begin{array}{cc} 
            \knl[(s)][t] & \knl[(c)][t] \\
            \knl[(c)][t] & \knl[(s)][t] \,
        \end{array}\right)
    \end{equation}
    where the kernel components have the exponential form
    \begin{equation}
        \knl[(s/c)][t]=\alp[(s/c)] \bet[(s/c)] e^{-\bet[(s/c)] t} \ind{t\in \mathbb{R}^+} \, ,
    \end{equation}
    where $\ind{x}$ is an indicator function equal to 1 if $x$ is true and zero otherwise. The above functions are  $\lone$-integrable, so that the choice $\alp[(s/c)],\bet[(s/c)] > 0$ ensures that the associated Hawkes process is well-defined.
    For the process to be stable, one needs to additionally require that the spectral norm satisfies  $|| \knl|| = || \knlmat[(s)]|| + ||\knlmat[(c)]|| < 1$. Due to $\alp[(s/c)] = \int_0^\infty \knl[(s/c)][t]$, the stability condition becomes
    \begin{equation}
        \alp[(s)] + \alp[(c)] < 1 \,.
    \end{equation}
    The $\alp[(s/c)]$ parameters can be interpreted as the ones setting the overall strength of the interactions, while the $\bet[(s/c)]$ control the relaxation time of the perturbations induced from past to future events.
\end{example}

The Hawkes process with exponential kernels has several advantageous properties, as it allows one to compute the expected value of arbitrary functions of $\cnt[t]$ (see Sec.~\ref{ssec:beyond_sec_ord} and Ref.~\cite{Errais:2010aa}), to be directly simulated (see App.~\ref{sec:simul}), or to compute efficiently its likelihood (see App.~\ref{ssec:param-estimation}). Most of these properties descend from a Markov property which in its simplest form is stated as follows:
\begin{prop}[Markov property for exponential kernels]
\label{prop:markov}
    Consider a Hawkes process with exponential kernels $\knl[ij][t] = \alp[ij] \beta e^{\bet t}\ind{t\in \mathbb{R}^+}$. Then the couple $(\cnt[t],\lam[t])$ is a Markov process. In particular, Eq.~(\ref{eq:hawkes}) for the intensity $\lam[t]$ can be recast in Markovian form as
    \begin{equation}
        \dd \lam[t] =  - \bet \lam[t] \, \dd t + \alp \bet \, \dd \cnt[t]
    \end{equation}
\end{prop}
This property can be extended to the case in which \emph{(i)} the coefficients $\bet[ij]$ are non-constant across components and \emph{(ii)} the kernel $\knl$ contains a finite sum of exponentials. The price to pay in this more general setting is the introduction of an extra set of $A$ auxiliary processes $\{\tilde \lam^{(a)}_t\}_{a=1}^A$, suitably chosen so that the resulting $(A+1)$-uple $(\cnt[t],  \tilde \lam^{(1)}_t,\dots,\tilde \lam^{(A)}_t )$ is Markovian.

In the non-exponential case, the Hawkes process cannot be generally mapped to a Markovian process, implying that it is necessary to take track of all its past history in order to perform exact simulation and estimation. A particularly well-known example of a non-exponential kernel is the power-law one, proposed in Ref.~\cite{Ogata:1988vn} in order to describe temporal clusters of seismic activity.
\begin{example}[Power-law kernel]
    Let's now consider the case $\nn=1$, and assume the kernel to be parameterized by the regularized power-law
    \begin{equation}
        \label{eq:knl_pow_law}
        \knl[t] = \frac{\alp \bet}{(1+\bet t)^{1+\gamma}} \ind{t\in \mathbb{R}^+}
    \end{equation}
    Also in this case the process is well-defined for $\alp,\bet > 0$. The stationarity condition in is met for
    \begin{equation}
        \alp < \gamma \,,
    \end{equation}
    indicating that the tail exponent $\gamma$ of a power law-kernel should be positive for the increments of the process to be stationary.

\end{example}

\subsection{Some extensions}
\label{ssec:extensions}

Although the model defined in above section is the one originally introduced in~\cite{Hawkes:1971lc,Hawkes:1971nq}, and most widely used in the literature, several generalizations have been proposed since.

\subsubsection{Marked Hawkes processes}
\label{ssec:marked}
Eq.~(\ref{eq:hawes_expl}) defining the Hawkes process can be enriched by endowing each event with a \emph{mark} variable, thus obtaining a sequence of event times, components and marks $\{ (t_m, k_m , \xi_m) \}_{m=1}^M$. One may further assumes events labeled with different marks to have different effects on the future intensities, leading to a dynamics for $\lam[t]$ of the type
\begin{equation}
    \lam[i][t] = \exo[i] + \sum_{m=1}^{M} \knl[i,k_m][t-t_m,\xi_m] \, ,
\end{equation}
Finally, one needs to introduce a generating mechanism for the marks, which are typically assumed to be i.i.d.\ random variables drawn with each event and sampled from a common distribution $p(\xi)$. A typical choice for the interaction kernel is the one $\knl[ij][t,\xi]=\knl[ij][t] \chi^{ij}(\xi)$, in which one assumes a factorized form for the effect of the marks.
This mechanism is used to describe events of different weights, and has been originally employed in order to model the occurrence of earthquakes of difference magnitudes~\cite{Ogata:1988vn}. In finance, marks can be used in order to model trades performed at times $t_m$ with different volumes $\xi_m$ (as e.g. in \cite{Fauth:2012,Bacry:2014c}, see Sec.~\ref{sec:price} and \ref{sec:impact}) or a drawdown intensity (as e.g. in \cite{Embrechts:2011aa,Chavez-Demoulin:2012aa}, see Sec.~\ref{ssec:risk}). On a more general ground, note that multivariate Hawkes processes can also be seen as an example of Hawkes processes with interacting marks~\cite{large_dim_hawkes}.

\subsubsection{Exogenous non-stationarity}
\label{ssec:exo_non_stat}
The exogenous intensity $\exo$ can be generalized to a deterministic function of time $\exotd[t]$. This choice allows to model a non-stationary system in which the interaction kernel is indeed independent of time. In finance, this is the case when one wants to model intra-day seasonalities (see the review article~\cite{Bauwens:2009aa}) and/or spillover effects within successive days (as in~\cite{Bowsher:2007aa}).

\subsubsection{Endogenous non-stationarity}
\label{ssec:endo_non_stat}
The scenarios $|| \Phi || > 1$ and $|| \Phi|| = 1$ have also been considered in order to model non-stationarity induced by endogenous interactions. These two cases present indeed an important difference: while in the former one the average intensity grows exponentially in time, in the latter one the process may possess a finite average event rate. This second type of non-stationarity, which we will call \emph{quasi-stationarity}, has raised a strong interest in the literature due to the fact that the condition $ || \Phi|| \approx 1$ is often met when calibrating Hawkes processes to real financial data (see the detailed discussion in Sec.~\ref{sec:univariate}). The limiting behavior of a Hawkes process in the regime $ || \Phi|| = 1$ has been analyzed by Br\'emaud and Massouli\'e in~\cite{Bremaud:2001aa}, where it is shown in particular that:
\begin{prop}[Degeneracy of critical, short-range Hawkes]
    Let $\cnt[t]$ be a univariate Hawkes process as in Eq.~(\ref{eq:hawkes}) such that $|| \Phi || = 1$ and $\exo=0$. Then if
    \begin{equation}
        \label{eq:short_range_knl}
        \int_0^\infty \dd t \, t \,\knl[t] < \infty \,
    \end{equation}
    the average of the conditional intensity is either $0$ or $+\infty$.
\end{prop}
Hence in the $\nn=1$ case, short-ranged kernels always lead to trivial processes. The next result shows instead that interactions of broader range allow instead a richer behavior.
\begin{theor}[Existence of critical, stationarity]
    Let's now consider a Hawkes process with $|| \Phi || = 1$, $\exo=0$ and
    \begin{eqnarray}
        \sup_{t\geq 0} t^{1+\gamma} \knl[t] &\leq& R \\
        \lim_{t\to\infty} t^{1+\gamma} \knl[t] &=& r
    \end{eqnarray}
    where $r,R >0$ and $\gamma \in \, ]0,1/2[$. Then the average intensity of such process is finite.
\end{theor}
Summarizing, there exist non-trivial univariate Hawkes processes with $||\knl||=1$ only for specific values of the tail exponent of $\knl[t]$, which is required to lie in the interval $\gamma \in \, ]0,1/2[$.

Notice that, even though in $\nn=1$ a quasi-stationary short-ranged Hawkes process is always degenerate, Ref.~\cite{Jaisson:2013aa} describes a scaling regime for $\cnt[t]$ in which it is possible to obtain to a non-degenerate process in the limit $||\knl||\to 1$ by appropriately choosing an observation timescale for the process. This behavior will be reviewed in Sec.~\ref{ssec:scaling-limit}.

In the multivariate setting, Ref.~\cite{large_dim_hawkes} shows that  even in presence of kernels $\knl$ satisfying the short-range condition Eq.~(\ref{eq:short_range_knl}) it is possible to define a non-trivial quasi-stationary Hawkes process in the large-dimensional limit. In particular Ref.~\cite{large_dim_hawkes} assumes a factorized form of the kernel, of the type
\begin{equation}
    \label{eq:factor_kernel}
    \knl[ij][t] = \alp[ij] f(t)\;,
\end{equation}
with $||f||=1$, and consider the $\nn\to\infty$ limit of the process $\cnt[t]$. Such a limiting regime turns out to be well-defined also when $||\knl|| =||\alp||\xrightarrow[\nn\to\infty]{} 1$, provided that the matrix $\alpha$ has a sufficiently low density of eigenvalues in the vicinity of the critical point $||\alp||=1$. Hence, a non-degenerate quasi-stationary limit for a Hawkes process can also be obtained as an effect of the interaction among a large number of components.

\subsubsection{Non-linear Hawkes}
\label{ssec:non_linear}
Non-linear generalizations of the Hawkes process have been considered by several authors \cite{Bremaud:1996aa,Sornette:2005,Rey0,Zheng:2014}. The intensity function in the non-linear case is written as
\begin{equation}
    \label{eq:non_lin_hawkes}
    \lam[i][t] = h \left( \exo[i] + \sum_{i=1}^{\nn} \int \dd\cnt[j][t'] \knl[ij][t-t'] \right) \, ,
\end{equation}
where $h(\cdot)$ is a non-linear function with support in $\mathbb{R}^+$. Typical choices for $h$ include $h(x) = \ind{x\in \mathbb{R}^+}$ and $h(x) = e^x$. Note that the stability condition for various functions $h$ was studied by Br\'emaud and Massouli\'e in \cite{Bremaud:1996aa}.
The main advantage introduced by this extension is the possibility of modeling inhibition through negative valued kernels, although the price that has to be paid is the loss of mathematical tractability for most of the properties of the process. Yet, simulation and calibration of the model are possible even in this scenario~\cite{Rey1,Bacry:2014ab}. As we will see, negative valued kernels are found in the context of finance (see for instance Sec.~\ref{ssec:level1}).

Let us end this section by mentioning that we have only referred to the most common extensions of Hawkes original model
but many further generalizations have been proposed like e.g., mixed diffusion-Hawkes models \cite{Errais:2010aa,Chavez-Demoulin:2012aa}, 
Hawkes models with shot noise exogenous events \cite{Dassios:2011}, Hawkes processes with generation dependent kernels \cite{Mehrdad:2014}.

\subsection{Properties}
\label{ssec:properties}
The linear structure of the stochastic intensity $\lam[t]$ of a Hawkes process allows us to characterize many of its properties in a completely analytical manner. Notably, its first- and second-order properties are particularly easy to compute, while a cluster representation of the process can be used in order to obtain a useful characterization of the Hawkes process. These properties are reviewed in the following Section.

\subsubsection{First and second order properties}
Assuming the hypothesis {\bf (H)}, it is then possible to write explicitly the first- and the second-order properties of the model in term of the Laplace transform of the kernel $\knl$. In order to do this, it is necessary to introduce the function $\ikn$ defined as follows:
\begin{defin}[Kernel inversion]
    Consider a  Hawkes process $\cnt[t]$ with stationary increments. We define $\ikn(t)$ as the causal solution of the equation
    \begin{equation}
        \label{eq:inv_kernel}
        \knl[t] + \ikn[t] * \knl[t] = \ikn[t]
        % \id \delta(t) = \left( \id \delta(t) + \ikn(t) \right) * \big( \id \delta(t) -\knl(t) \big)\, ,
    \end{equation}
    As a consequence of {\bf (H)}, $\ikn(t)$ exists and can be expressed as the infinite convolution
    \begin{equation}
        \label{eq:inf_conv}
        \ikn(t) = \knl(t) + \knl(t) * \knl(t) + \knl(t) * \knl(t) * \knl(t) + \dots
    \end{equation}
\end{defin}
The matrix function $\ikn$ can be characterized analytically in term of the Laplace transform of the kernel $\knl$, which we define as follows:
\begin{notat}[Laplace transform]
    Given a scalar function $f(t) \in L^1(-\infty,+\infty)$, we denote its Laplace transform as
        \begin{equation}
            \label{eq:laplace}
            \hat f(z) = \int_{-\infty}^\infty \dd t \, f(t) e^{zt}
        \end{equation}
    Vector and matrix Laplace transforms are defined by applying the above transformation component-wise.
\end{notat}
In the Laplace domain, Eq.~(\ref{eq:inf_conv}) is then mapped to the algebraic relation
\begin{equation}
    \label{eq:inv_kernel_laplace}
    \lik[z] = (\id - \lkn[z])^{-1} -\id \; ,
\end{equation}
where $\id$ denotes the identity matrix, allowing us to state the main result concerning the linear properties of a Hawkes process:
\begin{prop}[First- and second-order statistics]
    For a  Hawkes process $\cnt[t]$ with stationary increments, the following propositions hold:
    \begin{enumerate}
        \item The average intensity $\Lam = \av{\dd \cnt[t]} /\dd t$ is equal to
        \begin{equation}
            \label{eq:stat_lam}
            \Lam = (\id + \lik[0]) \mu
        \end{equation}
        \item The Laplace transform of the linear correlation matrix
        \begin{equation}
            \label{eq:corr}
            \cor[t-t'] = \frac{\av{\dd \cnt[t] \dd \cnt^T_{t'}}-\av{\dd \cnt[t] }\av{\dd \cnt^T_{t'}}}{\dd t \dd t'}
        \end{equation} is equal to
        \begin{equation}
            \label{eq:stat_corr}
            \lcor[z] = (\id + \lik[-z]) \, \Sig \, (\id + \lik^T(z) ) ,
        \end{equation}
        where $\Sigma$ is a diagonal matrix with non-zero elements equal to $\Sig[ii] = \Lam[i]$.
    \end{enumerate}
\end{prop} 
This useful characterization of the linear properties of a Hawkes process, first formulated in~\cite{Hawkes:1971lc,Hawkes:1971nq} and then fully generalized in~\cite{Bacry:2014ab}, allows to \emph{(i)} obtain the linear predictions of a Hawkes model given $\exo$ and $\knl$ as an input, \emph{(ii)} calibrate non-parametrically the kernel $\knl$ from empirical data by inverting relations~(\ref{eq:stat_lam}) and~(\ref{eq:stat_corr}) (see App.~\ref{ssec:param-estimation}).

Let us point out that, in Ref.~\cite{Bacry:2013aa}, the authors have established that under general conditions, the empirical covariation of a multivariate
Hawkes process converges towards its expected value, that can be easily expressed in terms 
of the Hawkes covariance matrix $\cor[t]$ as given by Eq.~(\ref{eq:stat_corr}).

\begin{figure}[htb]
    \begin{center}
        \includegraphics{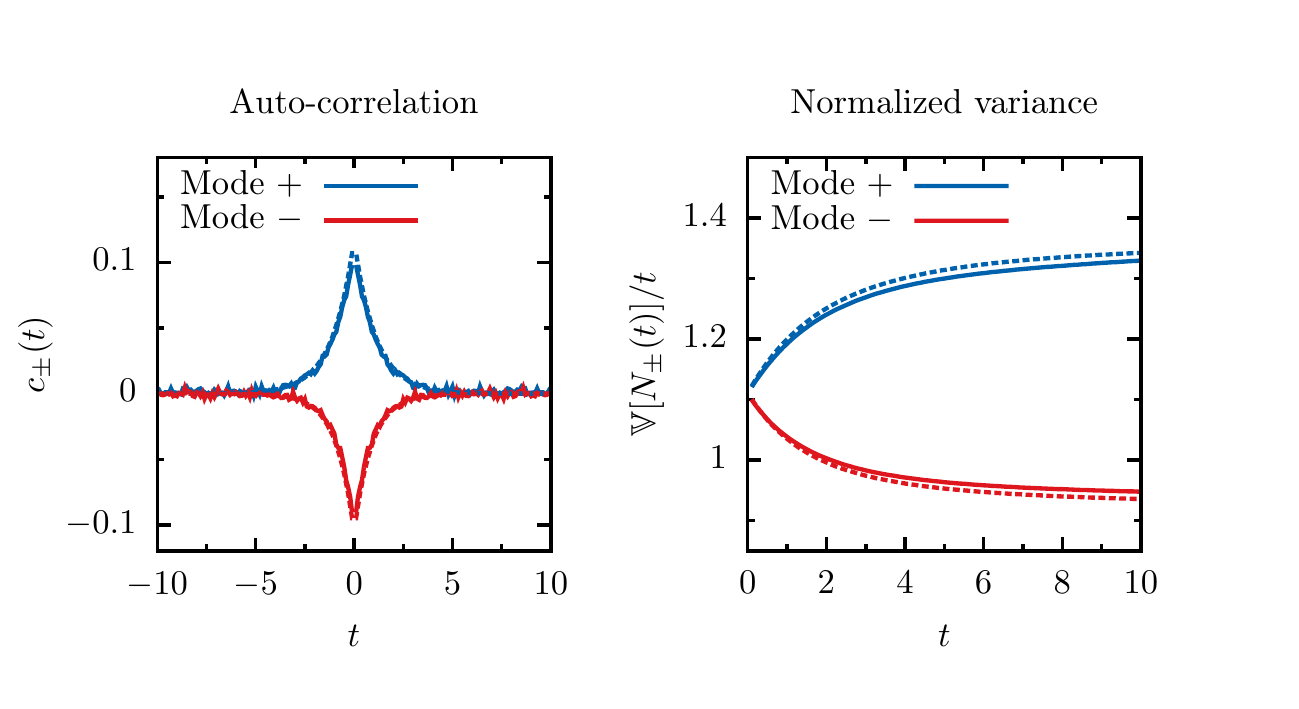}
    \end{center}
    \caption{Autocorrelation function (left) and normalized variance (right) of the combinations $\cnt[\pm][t]$ diagonalizing the interaction kernel appearing in Eq.~(\ref{eq:kernel_bivariate}). We have used the parameter set $\alp[(s)] = 0$, $\alp[(c)] = 0.1$, $\exo_0=\bet_0=1$ in order to simulate a single realization of the process of length $T=10^5$. For such a value of $T$, the theoretical predictions (dashed lines) are almost exactly superimposed to the results of the simulations. The $\delta(t)$ component in the cross-covariance function has been omitted in the left panel for the sake of clarity.}
    \label{fig:cor_cov}
\end{figure}

\begin{example}[Exponential kernel]
    Let's consider again the bivariate case described by Eq.~(\ref{eq:kernel_bivariate}) in the stationary case $ 1 > \alp[(s)] + \alp[(c)]$. In that case, the Laplace transform of the kernel $\lkn[z]$ reads:
    \begin{equation}
        \label{eq:kernel_bivariate_diag}
        \lkn[z] =  \frac{1}{2}\left(\begin{array}{cc}
            1 & 1 \\
            1 & -1 \,
        \end{array}\right)
        \left(\begin{array}{cc}
            \lkn[(s)][z] + \lkn[(c)][z] & 0 \\
            0 & \lkn[(s)][z] - \lkn[(c)][z] \,
        \end{array}\right)
        \left(\begin{array}{cc}
            1 & 1 \\
            1 & -1 \,
        \end{array}\right) \,,
    \end{equation}
    implying that the kernel matrix is diagonal in the symmetric and antisymmetric combinations $\cnt[\pm][t] = 2^{-1/2}(\cnt[1][t]\pm\cnt[2][t])$.
    The Laplace transform of the individual components of the kernel is
    \begin{equation}
        \label{eq:inv_exp_kernel_laplace}
    \lkn[z] = \frac{\alp[(s/c)]}{1-z/\bet[(s/c)]} \, ,
    \end{equation}
    and due to the diagonal form of Eq.~(\ref{eq:kernel_bivariate}) the inverse kernel $\lik[z]$ can be computed straightforwardly.
    If one assumes $ \exo=(\exo_0,\exo_0)$, then the relations above imply that $\Lam=(\Lam_0,\Lam_0)$, with
    \begin{equation}
        \label{eq:stat_int_exp}
        \Lam_0 = \frac{\exo_0}{1-\alp[(s)] -\alp[(c)]} \,.
    \end{equation}
    The Laplace transforms of the lagged cross-correlations of the symmetric and antisymmetric combinations $X_\pm(t)$, defined as
    \begin{equation}
        \label{eq:stat_cor_exp}
        \cor[\pm][t-t'] = \frac{\av{(\dd\cnt[1][t]\pm\dd\cnt[2][t'])(\dd\cnt[1][t']\pm\dd\cnt[2][t'])}}{2\dd t \dd t'}
        - \Lam_0^2 \,.
    \end{equation}
    result
    \begin{equation}
        \label{eq:stat_cor_exp_lapl}
        \lcor[\pm][z]= \frac{\Lam_0}{(1-\lkn[(s)][-z]\mp \lkn[(c)][-z])(1-\lkn[(s)][z]\mp \lkn[(c)][z])} \;.
    \end{equation}
    Above expression can be inverted explicitly so to obtain the lagged cross-correlations in real space. In the simpler case $\bet[(s)]=\bet[(c)]=\beta_0$, one obtains for example
    \begin{equation}
        \label{eq:stat_cor_exp_explicit}
        \cor[\pm][t] = \Lam_0 \left( \delta(t) + \frac{\beta_0}{2}\frac{(\alp[(s)] \pm \alp[(c)])(2-\alp[(s)] \mp \alp[(c)])}{(1-\alp[(s)] \mp \alp[(c)])} e^{-(1-\alp[(s)] \mp \alp[(c)])\beta_0 |t|}\right) \;.
    \end{equation}
    Note that a singular component arises for $t=0$ due to the assumption of unitary jumps $(\dd \cnt)^2 = \dd \cnt$. Above formula also shows that by moving the spectral norm $|| \knl||=\alp[(s)] + \alp[(c)]$ close to the instability point $||\knl||=1$, the decay of the symmetric mode of correlation function becomes slower and slower. Fig.~\ref{fig:cor_cov} illustrates the result above for the autocorrelation function of the modes $\cnt[\pm][t]$.
\end{example}

\begin{figure}[htb]
    \begin{center}
        \includegraphics{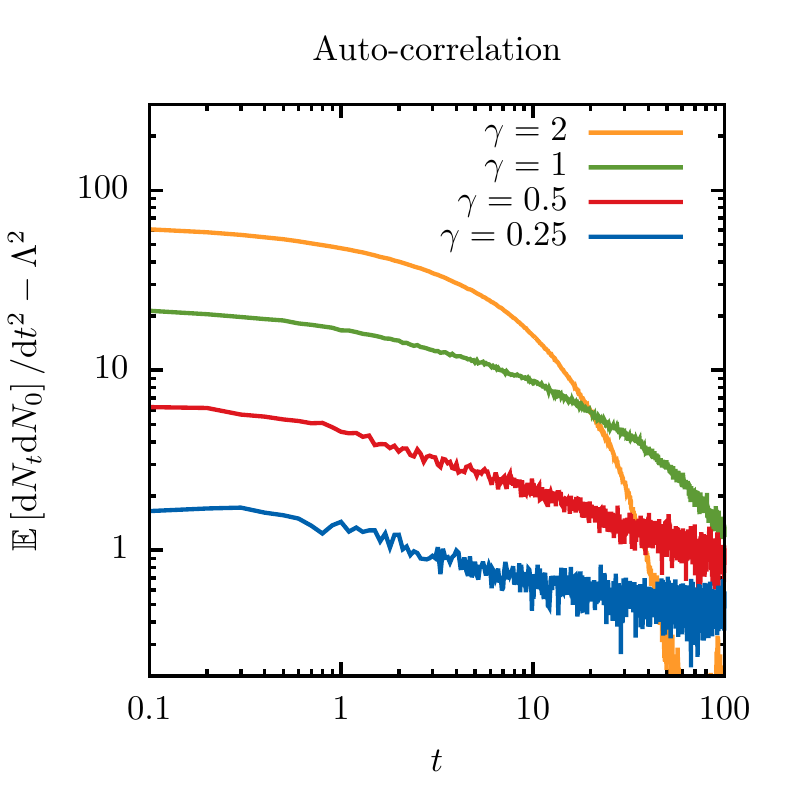}
    \end{center}
    \caption{Comparison of the autocorrelation function for a set of univariate Hawkes processes with power-law kernels parametrized by Eq.~(\ref{eq:knl_pow_law}). We used the values $\mu=\beta=1$, $\alpha = 0.9\,\gamma$, and simulated processes of length  $T=5 \cdot 10^5$ (for $\gamma=0.25$ and $0.5$), $T=10^6$ (for $\gamma=1$) and  $T=5\cdot 10^6$ (for $\gamma=2$) in order to obtain the curves represented in the plot. The image illustrates the crossover from the exponential decay of correlation obtained for $\beta>1$ to the power-law behavior detected for $\beta<1$.}
    \label{fig:cor_pow_law}
\end{figure}

\begin{example}[Power-law kernel]
Let's now consider again the case of a power-law interaction kernel in dimension $\nn=1$. For a kernel parameterized by Eq.~(\ref{eq:knl_pow_law}), the Laplace transform reads
\begin{equation}
    \label{eq:laplace_pow_law}
            \lkn[z] = \alpha e^{-z/\beta} (-z/\beta)^{\gamma} \Gamma(-\gamma,-z/\beta) \, ,
\end{equation}
where $\Gamma(n,z)$ is the incomplete Gamma function. The inverse kernel $\lik[t]$ may be expressed in the Laplace domain as
\begin{equation}
    \label{eq:psi_exp}
    \lik[z] = \frac{\alpha e^{-z/\beta} (-z/\beta)^{\gamma} \Gamma(-\gamma,-z/\beta)}{1-\alpha e^{-z/\beta} (-z/\beta)^{\gamma} \Gamma(-\gamma,-z/\beta)} \, .
\end{equation}
As shown in Sec.~\ref{ssec:defin}, in this case the spectral radius $||\knl ||$ is equal to $||\knl || = |\lkn[0]| = \alpha/\gamma$, so that the model is stable for $\alpha < \gamma$. Under this assumption, the average intensity results
\begin{equation}
    \label{eq:stat_int_pow_law}
    \Lam = \exo \left( \frac{\gamma}{\gamma-\alpha} \right) \, .
\end{equation}
For a fixed value of the exogenous intensity $\exo$, this relation interpolates between a total intensity equal to the exogenous one (in the non-interacting case $\alpha=0$) and an increasingly larger number of events as soon as $\alpha$ approaches the instability point $\alpha=\gamma$.\\
The Laplace transform of the lagged cross-correlation matrix results
\begin{equation}
    \label{eq:stat_cor_lapl_pow_law}
    \lcor[z] = \frac{\Lam}{
    (1- \lkn[-z])(1- \lkn[z])
    }\;.
\end{equation}
The above expression cannot be inverted analytically. One can indeed relate the tail behavior of the correlations to the small $z$ behavior of $\lcor[z]$ thanks to Tauberian Theorems. In particular, for $\beta t \gg 1$ one has:
\begin{equation}
    \label{eq:asymp_pow_law}
    \cor[t] \sim
    \left\{
    \begin{array}{lcr}
    e^{ - (\gamma-1) \beta t} & \textrm{for} & \gamma > 1 \\
    (\beta t)^{-\gamma -1} & \textrm{for} & \gamma < 1
    \end{array}
    \right.
\end{equation}
This behavior is illustrated in Fig.~\ref{fig:cor_pow_law}, where we compare the autocorrelation function of several univariate Hawkes processes with power-law kernel and different tail exponents.
As a final note, we remark that for $\gamma<1/2$, and close to the instability point $\alpha = \gamma$, the function $\cor[t]$ obeys an intermediate asymptotics $\cor[t] \sim t^{2\gamma-1}$, which holds as long as
\begin{equation}
    \label{eq:approx_criticality}
    \beta t \ll \left( \frac{\Gamma(1-\gamma)}{\gamma/\alpha -1}\right)^{1/\gamma} \, .
\end{equation}
In this particular regime, the Hawkes process develops an apparent Hurst exponent $H=1/2+\gamma$. This limiting behavior is a  consequence of the quasi-stationarity condition $||\knl||=1$, analyzed in~Ref.~\cite{Bremaud:2001aa} and reviewed in Sec.~\ref{ssec:endo_non_stat}.
\end{example}

\subsubsection{Characterization through second order statistics} % (fold)
\label{ssub:characterization_through_second_order_statistics}

This section justifies why the Hawkes process can be thought of as the simplest example of an interacting point process. In fact, it is entirely characterized by its first- and second-order properties: means and correlations uniquely determine a Hawkes process through the solution of a Wiener-Hopf system.

Let us start by defining the \emph{conditional intensity} matrix $\cnd[t]$, that we define for $t>0$ as
\begin{equation}
     \label{eq:cond_int}
     \cnd[ij][t] = \frac{\avc{\dd \cnt[i][t]}{\dd \cnt[j][0]=1}}{\dd t} -\Lam[i] \;, ~~~\forall t >0.
 \end{equation}
 Then one can prove straightforwardly \cite{Bacry:2014ab} from Eq.~(\ref{eq:stat_corr}) that
 \begin{equation}
     \label{eq:cor_vs_cnd}
     \cor[t] = \Sig \cnd^T(t) \,,~~~ \forall t >0
 \end{equation}
 which relates conditional averages and lagged cross-correlations. Hence by using Eq.~(\ref{eq:cor_vs_cnd}) and (\ref{eq:stat_corr}), one can prove that \cite{Bacry:2014ab}:
\begin{theor}[Wiener-Hopf equation]
    Consider a Hawkes process defined by Eq.~(\ref{fig:hawkes}) satisfying the stationarity assumption {\bf (H)}. Then the matrix function $\chi(t) = \knl[t]$ is the unique solution of the Wiener-Hopf system
    \begin{equation}
        \label{eq:wiener-hopf}
        \cnd[t] = \chi(t) + \chi(t) * \cnd[t] \quad \forall t > 0
    \end{equation}
    such that the components $\chi^{ij}(t)$ are causal and $ \chi^{ij}(t) \in \lone \; \forall i,j$.
\end{theor}
This property implies that, when fixing an average intensity vector $\Lam$ and a conditional expectation $\cnd[t]$, there exist at most one Hawkes process consistent with these observables. Indeed, such a process is not always guaranteed to exist, as a Hawkes process doesn't necessarily reproduce the linear properties for systems in which inhibition is relevant.

This result is the inverse one with respect to the one expressed by Eq.~(\ref{eq:stat_corr}): while that equation expresses the fact that by fixing a kernel $\knl[t]$ and an exogenous intensity $\exo$ the correlations are uniquely determined, the theorem above states that correlations and average intensities uniquely fix the interactions. While the direct result was first proved in~\cite{Hawkes:1971lc}, the converse was shown in~\cite{Bacry:2014ab} by using the Wiener-Hopf factorization technique. 

Finally note that the Wiener-Hopf system~(\ref{eq:wiener-hopf}) is very useful in applications to empirical data, as it allows us to estimate non-parametrically the interaction kernel of a Hawkes process given a set of empirical observations (see App.~\ref{ssec:nonparam}).

% subsubsection characterization_through_second_order_statistics (end)

\subsubsection{Auto-regressive projection}
A Hawkes process with stationary increments can always be linearly approximated by suitably defined auto-regressive processes. In particular one can show that \cite{large_dim_hawkes}:
\begin{prop}[Auto-regressive projection]
    Consider a Hawkes process $\cnt[t]$ defined by Eq.~(\ref{eq:hawkes}) satisfying the stationarity assumption {\bf (H)}. Then the convolution $\cnt^{(AR)}_t$ defined by
    \begin{equation}
        \label{eq:ar}
        \cnt^{(AR)}_t = \left( \id\delta(t) + \ikn[t] \right) * (\exo \, t + \Sig^{1/2} W_t) \;,
    \end{equation}
    where $\ikn[t]$ is the infinite convolution of the Hawkes kernel given by Eq.~(\ref{eq:inf_conv}), and $W_t$ is a standard $\nn$-dimensional Brownian motion, satisfies
    \begin{eqnarray}
        \label{eq:ar_matching}
        \Lam^{(AR)} &=& \Lam \\
        \cor^{(AR)}(t) & = & \cor[t] \, ,
    \end{eqnarray}
    where $\Lam$ and $\cor$ denote respectively the average intensity and the lagged cross-correlation matrix of $\cnt[t]$.
\end{prop}
Hence, it is always possible to match the first and the second order properties of a stationary Hawkes process with interaction kernel $\knl$ by using a convolution of Wiener processes. On the other hand, higher order moments cannot be matched in the same way.

\subsubsection{Beyond second-order}
\label{ssec:beyond_sec_ord}
The first- and second-order moments are not the only moments which can be computed analytically for a Hawkes process. In particular, Jovanovi\'c \etal~\cite{Jovanovic:2014aa} have recently developed a combinatorial procedure allowing to calculate cumulants (and consequently, moments) of arbitrary order of a Hawkes process. More precisely, given a set of components $S \in \{1,\dots,\nn\}$ and one of times $t_S=\{ t_1,\dots t_{|S|}\}$, it is possible to define a \emph{cumulant density} of a Hawkes process as
\begin{equation}
    \label{eq:cumulants}
    k\left(\cnt^{(S)}\right) =  \dd t^{-|S|} \sum_\pi (|\pi|-1!) (-1)^{|\pi|-1}\prod_{B \in \pi} \left< \prod_{i \in B} \dd \cnt[i][t_i] \right> \, ,
\end{equation}
where the sum runs over all the partitions $\pi$ of $S$, $|\pi|$ denotes their number of blocks and $B$ labels individually the blocks of $\pi$. Above equation generalizes the definition of the average intensity, recovered for $|S|=1$, and of the lagged cross-correlation function, obtained for $|S|=2$. Moment densities can be obtained from above expression by writing
\begin{equation}
    \label{eq:moments}
    \left< \prod_{i \in S} \dd \cnt[i][t_i] \right> \dd t^{-|S|} = \sum_\pi \prod_{B \in \pi} k\left(\cnt^{(B)}\right) \,.
\end{equation}
Ref.~\cite{Jovanovic:2014aa} shows how to express Eq.~(\ref{eq:cumulants}) as a sum of integral terms, which can be written explicitly in terms of $\exo$ and $\ikn[t]$. As each of such addends can be interpreted as a topologically distinct rooted tree with $|S|$ labeled leaves, the enumeration of all the contribution to Eq.~(\ref{eq:cumulants}) can be performed systematically.

In the special case of a Hawkes processes with exponential kernel, a useful result is obtained by Errais \etal~\cite{Errais:2010aa} by exploiting Dynkin's formula for the couple $(\cnt,\lam)$ in the marked framework described in Sec.~\ref{ssec:marked}. In particular, they are able to express the generating function for the couple $(\cnt,\lam)$ in terms of the solution of an ordinary differential equation. While it may be necessary to solve numerically the equation for the generating function, closed-form expressions are available for specific moments of $\cnt[t]$ (see also the work of Dassios and Zhao \cite{Dassios:2011} for
similar results on a slight generalization of Hawkes processes).

\subsubsection{Martingale representation}
The process defined by Eq.~(\ref{eq:hawkes}) admits a convenient martingale representation once one introduces suitably defined \emph{compensators} $\int_0^t \dd s\, \lam[i][s]$ \cite{daley1988introduction}.
\begin{theor}[Martingale representation]
    Given a Hawkes process~(\ref{eq:hawkes}), the $\nn$ stochastic processes
    \begin{equation}
        \label{eq:martingale_count}
        \mart[t] = \cnt[t] - \int_0^t \dd s\, \lam[s]
    \end{equation}
    are martingales with respect to the canonical filtration of the process $\cnt[t]$~\cite{daley1988introduction}. Additionally, under the stability condition {\bf (H)}, the stochastic intensity $\lam[t]$ admits the representation
    \begin{equation}
        \label{eq:martingale_intensity}
        \lam[t] = \exo + \int_0^t \ikn[t-s] \mu \, \dd s  + \int_0^t \ikn[t-s] \, \dd \mart[s] \, .
    \end{equation}
\end{theor}
While the above result is valid even in the non-stationary regime, Eq.~(\ref{eq:martingale_intensity}) takes a particularly simple form in the asymptotic regime of large $t$, where it can be written as
\begin{equation}
    \label{eq:martingale_stat}
    \lam[t] \xrightarrow[t\to\infty]{} \Lam  + \int_0^t \ikn[t-s] \, \dd \mart[s]
\end{equation}
thanks to Eq.~(\ref{eq:stat_lam}).
The martingale property of the Hawkes process plays an important role in determining the first- and second-order properties discussed in above section, which are derived by means of Eq.~(\ref{eq:martingale_stat}) in \cite{Bacry:2011kx,Bacry:2014ab}. The representation Eq.~(\ref{eq:martingale_intensity}) is also particularly useful in order perform predictions of the intensity of the Hawkes process given an historic filtration $\filt{t}$ of the process.
\begin{example}[Prediction of the intensity]
Suppose that, given a Hawkes process which satisfies the stationary condition {\bf (H)}, one is interested in computing the predictor $\avc{\lam[t]}{\filt{s}}$ for $t>s$.
 (see Ref.~\cite{Hewlett:2006aa,Jaisson:2014aa} for direct applications in Finance). While, by naively applying the definition of
 the Hawkes processes, one can obtain an implicit equation of the type
\begin{equation}
     \label{eq:pred_implicit}
     \avc{\lam[t]}{\filt{s}} = \exo + \int_0^s  \knl[t-u] \dd\cnt[u]
     + \int_s^t \dd u \, \knl[t-u] \avc{\lam[u]}{\filt{s}} \,
 \end{equation} 
 the martingale representation Eq.~(\ref{eq:martingale_intensity}) can be used in order to write the explicit expression
 \begin{equation}
    \label{eq:pred_expl}
     \avc{\lam[t]}{\filt{s}} = \exo + \int_0^t \ikn[t-s] \mu \, \dd s  + \int_0^s \ikn[t-s] \, \dd \mart[s] \, . \,
 \end{equation}     
\end{example}

\subsubsection{Scaling limit of the process}
\label{ssec:scaling-limit}
The structure of Hawkes processes is naturally adapted to describe systems in which the discrete nature of the jumps in the coordinates $\cnt[t]$ is relevant, making this model especially suitable for modeling high-frequency data. Indeed, in many applications one additionally needs to control the limiting behavior the system in the opposite regime of low frequencies, where the granularity of events is disregarded, and the scaling in time of $\cnt[t]$ needs to be known. 

\begin{figure}[htbp]
    \begin{center}
        \includegraphics{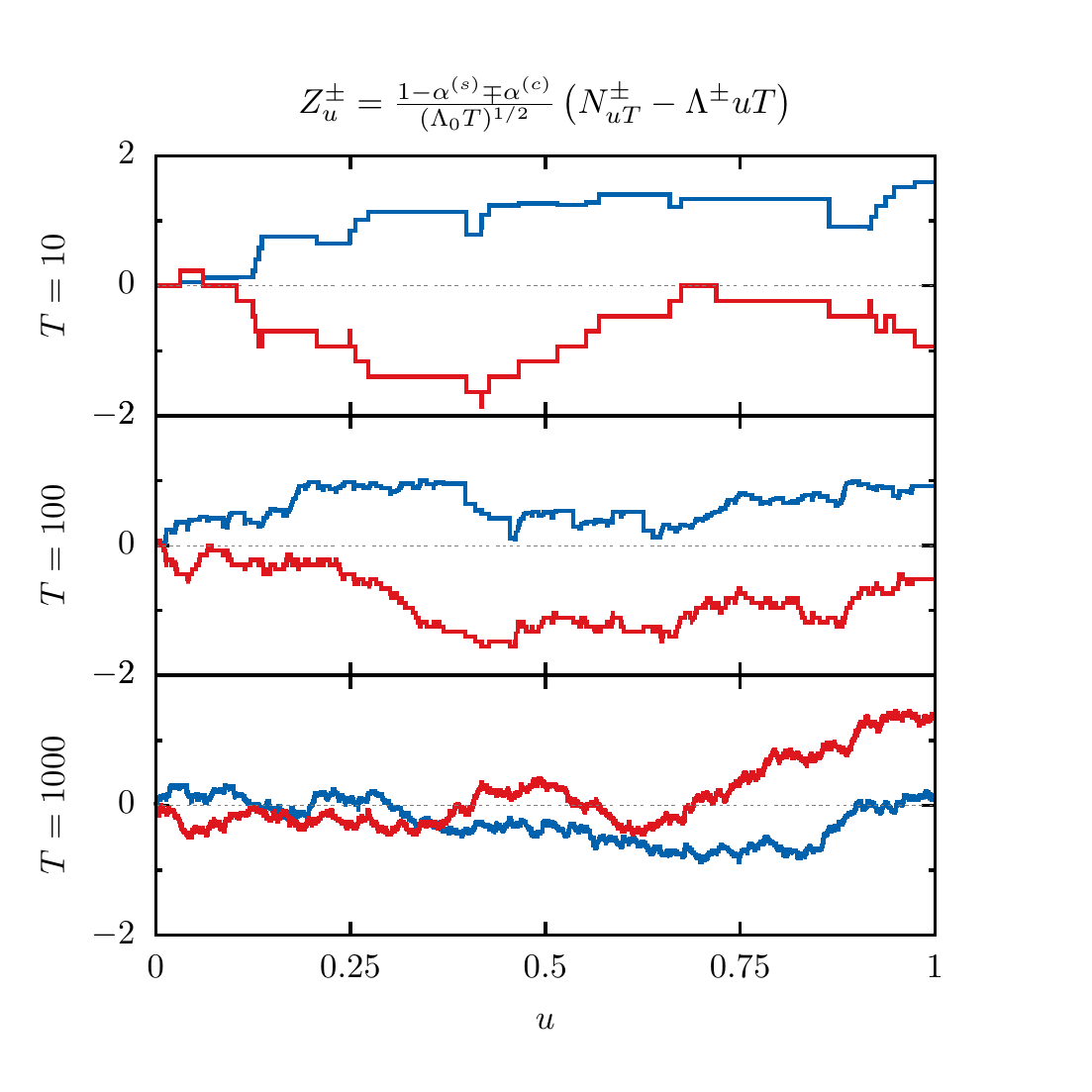}
    \end{center}
    \caption{The figure illustrates the shape of the process $\cnt[\pm][uT]$ at different timescales $T=10,100,1000$ for the same choice of parameters as in Fig.~\ref{fig:cor_cov}. The process has been rescaled according to Eq.~(\ref{eq:ex1diffusive}) so to obtain a standard Wiener process $\wien[u]$ in the limit $T\to\infty$. The plot illustrates how the presence of discrete jumps, clarly visible at small times, becomes irrelevant in the scaling limit $T\to\infty$.}
    \label{fig:diff_lim}
\end{figure}

\paragraph{Diffusion towards a Brownian motion.}
The following results, first proved in~\cite{Bacry:2013aa} allow us to establish that Hawkes processes, under appropriate hypotheses and after a suitable rescaling, behave at large times as linear combinations of Wiener processes.
\begin{theor}[Law of large numbers]
    Consider a Hawkes process as in Eq.~(\ref{eq:hawkes}) satisfying the stationarity assumption {\bf (H)}. Then
    \begin{equation}
        \label{eq:law_large_numb}
        \sup_{u \in [0,1]} || T^{-1} \cnt[uT] - u \Lam || \xrightarrow[T\to\infty]{} 0
    \end{equation}
    almost surely and in $L^2$-norm.
\end{theor}
Above result establishes a law of large numbers for the Hawkes process, valid for any stationary kernel. Indeed, under additional hypotheses it is also possible to formulate a corresponding functional central-limit theorem.
\begin{theor}[Central-limit theorem]
\label{th:limit}
    Suppose that for all $i,j \leq \nn$ the kernel $\knl[t]$ satisfies
    \begin{equation}
        \label{eq:centr_lim_theor}
        \int_0^\infty \dd t \, t^{1/2} \knl[ij][t] < \infty \;.
    \end{equation}
    Then for $u \in [0,1]$ one has the following convergence is in law for the Skorokhod topology:
    \begin{equation}
        \label{eq:scal_lim}
        T^{1/2} \left( T^{-1} \cnt[uT] - u\Lambda \right) \xrightarrow[T\to\infty]{}
        (\id + ||\ikn||) \Sig^{1/2} \wien[u] \, ,
    \end{equation}
    where $\wien[t]$ denotes a standard $\nn$-dimensional Brownian motion.
\end{theor}
\begin{example}[Exponential kernel]
    Consider the bivariate Hawkes process analyzed in Sec.~\ref{ssec:defin}. Eqs.~(\ref{eq:kernel_bivariate}) and~(\ref{eq:scal_lim}) above imply that
    \begin{eqnarray}
        \label{eq:scal_lim_exp}
        \cnt[uT] &\xrightarrow[T\to\infty]{}&
        \Lam_0 uT \left(
        \begin{array}{c}
            1 \\ 1
        \end{array}\right) \\
        &+& 
        \frac{(\Lam_0 T)^{1/2}}{(1-\alp[(s)])^2 -(\alp[(c)])^2}\left(
        \begin{array}{cc}
            1-\alp[(s)] & \alp[(c)] \\
            \alp[(c)] & 1-\alp[(s)]
        \end{array}\right)
        \left(
        \begin{array}{c}
            \wien[1][u] \\
            \wien[2][u]
        \end{array}
        \right)
        \,, \nonumber
    \end{eqnarray}
    which in term of the combinations $\cnt[\pm][t]=2^{-1/2} (\cnt[1][t]\pm \cnt[2][t])$ reads
    \begin{equation}
    \label{eq:ex1diffusive}
        \cnt[\pm][uT] \xrightarrow[T\to\infty]{} \left( 1\pm 1 \right)^{1/2} \Lam_0 u T +
        \frac{(\Lam_0 T)^{1/2}}{1-\alp[(s)] \mp \alp[(c)]} \wien[\pm][u] \, .
    \end{equation}
    This behavior is summarized in Fig.~\ref{fig:diff_lim}, where we compare the rescaled processes $\cnt[\pm][uT]$ at different timescales $T$.
\end{example}
\paragraph{Other scaling limits}
It is well known that many of the microscopic observables involved in the price formation process do not diffuse at large scales (e.g., at the daily, or even monthly, time-scale) toward Brownian motion dynamics. For instance, the trading activity (i.e., the market-order flow) seems to be 
long-range dependent \cite{BouchaudFarmerLillo},
while the volatility displays clusters that can last for months. Since, as we will see in the next sections, Hawkes processes tend to mimic very well the high-frequency dynamics of financial time-series, it is natural to try to understand if diffusive limits other than the one described by Theorem~\ref{th:limit} can be found. 
In a recent work by Jaisson and Rosenbaum~\cite{Jaisson:2013aa}, a sequence of rescaled univariate Hawkes processes
\begin{equation}
    \label{eq:scal_lim_cir}
    Z^{(T)}_u = \frac{1-a_T}{T}\cnt^{(T)}_{uT}
\end{equation}
indexed by a time-scale parameter $T$ and where $a_T \in [0,1[$ is proved to converge, when $T\rightarrow +\infty$, towards an integrated Cox, Ingersoll, Ross process~\cite{CIR:1985} under the main following conditions \footnote{For precise formulation of the corresponding theorem, we refer the reader to~\cite{Jaisson:2013aa}.}
\begin{itemize}
    \item the corresponding sequence of kernels $\knl[(T)][t]$ satisfies $\knl[(T)][t] = {a_T} \knlmat(t)$, where  $\knlmat$ is differentiable and it is such that $||\knlmat|| = \hat \phi(0) = 1$, $\hat \phi'(0) < +\infty$,
    %$\int_0^{+\infty} \phi(s)\dd s = 1$, $\int_0^{+\infty} s\phi(s)\dd s < +\infty$, 
    $||\phi'||_\infty < +\infty$
    and $||\phi'|| < +\infty$,
    \item the criticality condition {\bf (H)} of Prop.~\ref{prop:H} is met at a speed
    \begin{equation}
        \label{eq:scaling_lim_norm}
        \lim_{T\rightarrow +\infty} (1-a_T) T = \kappa,~~~~\kappa>0.
    \end{equation}
    \end{itemize}
Hence, even though a quasi-stationary short-ranged Hawkes process is always degenerate, one can detect a non-trivial behavior of the rescaled counting function by suitably choosing an observation timescale $T\sim (1-||\knl^{(T)}||)^{-1}$.
Let us point out that the above conditions do not allow for $\knlmat$ to have a power-law decay with an exponent strictly smaller than 2 (in the limit $t\rightarrow +\infty$). As reviewed in Sec.~\ref{sec:univariate}, several studies tend to show that for many financial time-series (e.g., market order flow time-series) the relevant kernel has power-law tails with an exponent above but rather close to 1. Thus, strictly speaking, the previous framework is inappropriate to describe such behavior.
Using the same asymptotics ($T\rightarrow +\infty)$, Jaisson~\cite{Jaisson:2014aa} studied the asymptotic limit of the correlation function of a 1-dimensional Hawkes process with a power-law kernel which decreases with an exponent $1+\gamma$ with $\gamma \in \, ]0,1/2[$. He proved that the auto-correlation function of the Hawkes process decreases asymptotically as a power-law with an exponent $1-2\gamma$, i.e., leading to a long-range dependence\footnote{Let us notice that qualitative arguments for similar result were also given in \cite{Bacry:2014aa} in a 2-dimensional framework.}.

\subsubsection{Clustering representation}
\label{ssec:cluster_repr}
\begin{figure}[htb]
    \begin{center}
        \includegraphics{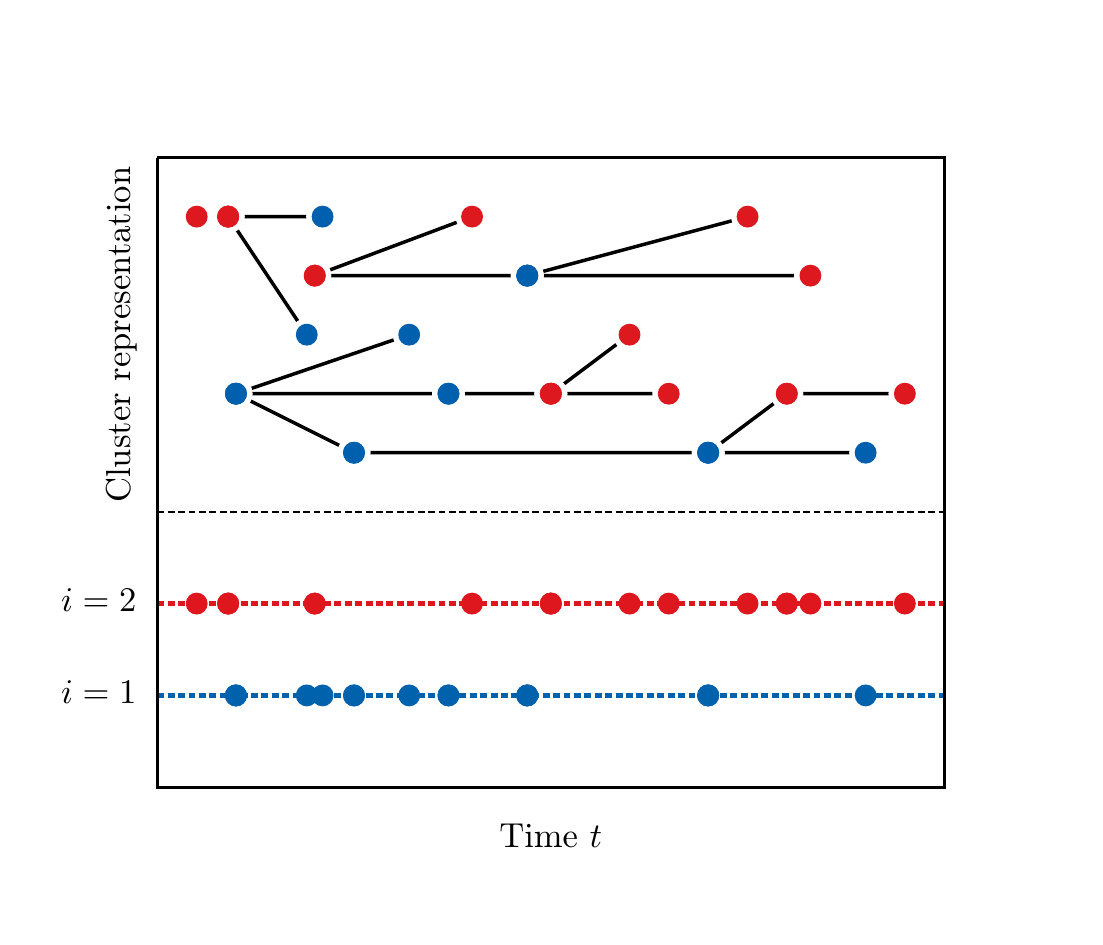}
    \end{center}
    \caption{Cluster representation of a Hawkes process: while the upper panel represents the branching structure of a bivariate Hawkes process, the lower panel shows its projection obtained by disregarding the cluster structure. The different components $i\in \{1,2\}$ are shown in different colors, while the connected structures in the upper panel denote three different clusters.}
    \label{fig:cluster}
\end{figure}

Another useful property of the Hawkes process is the \emph{clustering} property which emerges as a consequence of the linearity of Eq.~(\ref{eq:hawkes}). Such property allows one to \emph{(i)} build an efficient \emph{simulation} algorithm for the process (see App.~\ref{sec:simul}), \emph{(ii)} introduce the notion of \emph{parenthood} among different events (see Sec.~\ref{ssec:causality} below), \emph{(iii)} \emph{infer} parenthood relations among successive events from empirical data (see the paragraph about the EM method in App.~\ref{ssec:nonparam}). 
\begin{prop}[Clustering representation]
    Consider a positive integer $\nn$ and a (non-necessarily finite) time interval $[0,T]$, in which we define a sequence of \emph{events} $\{ (t_m,k_m) \}_{m=1}^M$ according to the following procedure:
    \begin{itemize}
        \item For each $1\leq i \leq \nn$, consider a set of \emph{immigrant} events $\{ (t_m^{(0)},i) \}_{m=1}^{M^{(0)}_i}$ extracted with homogeneous Poissonian rate $\exo[i]$ in the interval $[0,T]$.
        \item For each immigrant event of type $j$, labeled by $ (t_{m'}^{(0)},j)$, and for each $1\leq i \leq N$, generate a sequence of \emph{first-generation} events $\{(t^{(1)}_{m},i)\}_{m=1}^{M^{(1)}_i}$ sampled with time-dependent Poissonian rate $\knl[ij][t-t_{m'}]$ in the interval $[t_{m'}^{(0)},T]$.
        \item Iterate above rule from generation $n-1$ to generation $n$, so to obtain the event sequence $\{ (t^{(n)}_m,k_m^{(n)})\}_{m=1}^{M^{(n)}}$, until no more events are generated in $[0,T]$.
    \end{itemize}
    Then the union of all the events
    \begin{equation}
        \label{eq:cluster_union}
       \{ (t_m,k_m) \}_{m=1}^M = \bigcup_{n=0}^\infty \{ (t_m^{(n)},k_m^{(n)}) \}_{m=1}^{M^{(n)}}
    \end{equation}
    corresponds to the one generated by the Hawkes process~(\ref{eq:hawkes}) in the time interval $[0,T]$.
\end{prop}
Note that the construction above (depicted in Fig.~\ref{fig:cluster}) can be equivalently taken as a definition for the Hawkes process, once the information encoding the generation $n$ is discarded by taking the union of all the events.
Indeed, this richer definition characterizes more transparently the stationarity condition {\bf (H)}: by considering $T=\infty$ in above construction, it is possible to map the branching structure of the Hawkes process described above onto the one of a Galton-Watson tree with average offspring $||\knl||$. The qualitative behavior of the model is in fact dictated by the average number of events generated by a parent event, equal to $\int_0^\infty \knlmat(t) = \lknmat(0) = || \knl ||$. The three phases of the Hawkes process (stationary $||\knl|| < 1 $, non-stationary $|| \knl || > 1$ and quasi-stationary $||\knl||=1$) correspond then to the three phases of a Galton-Watson branching process, more precisely:
\begin{enumerate}
    \item For $||\knl|| < 1 $, we have a \emph{sub-critical} phase in which each parent event generates on average less than one child event. This implies that the total progeny of each event is a.s.\ finite and  the average number of generations before extinction is a.s.\ finite.
    \item For $||\knl|| > 1 $, we have a \emph{super-critical} phase in which more than one child event is generated by each parent event. In that case the total progeny of a parent event might be infinite with finite probability.
    \item For $||\knl|| =1$ (the \emph{critical} case), the total progeny is a.s.\ finite, but the total size of the progeny has large fluctuations leading to a divergence of the average number of generations before extinction.
\end{enumerate}

\subsubsection{Causality} % (fold)
\label{ssec:causality}
The parenthood relation introduced by using the clustering representation of the Hawkes process allows us to discuss the problem of causality in this context. In particular, after constructing a Hawkes process according to the branching procedure described above, one can introduce the counting functions
\begin{eqnarray}
    \label{eq:partial_cnt}
    \cnt[i\leftarrow 0][t] &=& (\textrm{Exogenously generated events of type } i) \\
    \cnt[i\leftarrow j][t] &=& (\textrm{Events of type } i \textrm{ with type } j \textrm{ direct ancestor}) \\
    \cnt[i\leftarrow j*][t] &=& (\textrm{Events of type } i \textrm{ with type } j \textrm{ oldest ancestor})
\end{eqnarray}
so that $\cnt[i\leftarrow 0][t] + \sum_j \cnt[i\leftarrow j][t] = \cnt[i\leftarrow 0][t] + \sum_j \cnt[i\leftarrow j*][t] = \cnt[i][t]$. Hence, these quantities can be used in order to express the overall number of events of type $i$ generated by an ancestor of a given type $j$. In particular, it is easy to prove that:
\begin{prop}[Causality]
    For a stationary Hawkes process, the average increments of $\cnt[i\leftarrow 0][t]$, $\cnt[i\leftarrow j][t]$ and $\cnt[i\leftarrow j*][t]$ are expressed by
    \begin{eqnarray}
        \label{eq:avg_part_cnt}
        \av{\dd \cnt[i\leftarrow 0][t]}/ \dd t & = & \exo[i] \\
        \av{\dd \cnt[i\leftarrow j][t]}/ \dd t  & = & \lkn[ij][0]\Lam[j] \\
        \av{\dd \cnt[i\leftarrow j*][t]}/ \dd t  & = & \lik[ij][0]\exo[j] \; .
    \end{eqnarray}
\end{prop}
The property above can be used in order to estimate the average fraction of events (directly or indirectly) caused by a specific component of a Hawkes process.

Alternative notions of causality for point-processes have also been investigated. In particular, the notion of Granger-causality has been extended to point-processes in~\cite{Nedungadi:2009kx,
Dhamala:2008vn}.

% subsection causality (end)

\section{Univariate models}
\label{sec:univariate}
\subsection{Models of market activity and risk}
\label{ssec:risk}
The first straightforward application of Hawkes processes in
high frequency finance is probably to model the so-called volatility clustering
phenomenon. Since volatility at the transaction level can be directly related
to the number $\cnt[T]$ of a given type of events (trades, mid-price changes,...)
that occur in a given time interval of size $T$\footnote{One can have in mind a simple price model 
where the price is build as the sum of independent random shocks. In this case, the volatility
at scale $T$ is proportional to $(\cnt[T])^{1/2}$. The empirical results of Ref.~\cite{wyart2008relation} corroborate empirically this observation, allowing one to establish more precisely of the relation among impact per trade and volatility.}, the self-exciting nature of Hawkes processes
provides a very simple picture that can explain 
the correlated nature of volatility fluctuations.
This idea was first proposed by Bowsher~\cite{Bowsher:2007aa} who calibrated a univariate
Hawkes model with mixture of exponential kernels 
using intraday equity data from NASDAQ and NYSE~\footnote{More precisely Bowsher considered a generalization of Hawkes processes in order to account for the peculiar non-stationarities observed at high frequency like intraday seasonalities and overnight gaps.}.

\begin{figure}[htb]
    \begin{center}
        \includegraphics{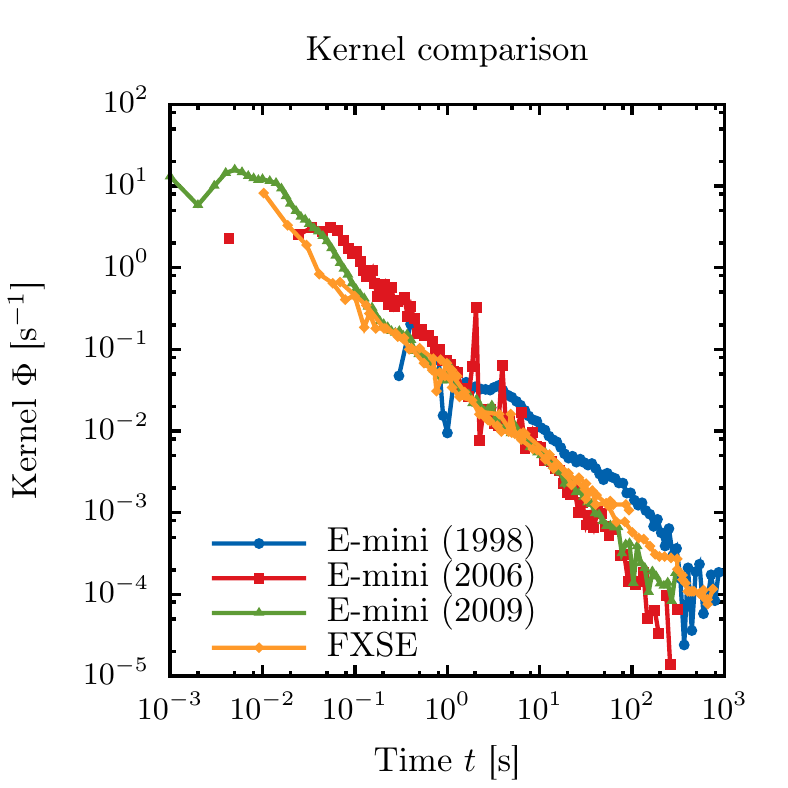}
    \end{center}
    \caption{Inferred kernel $\knl[t]$ of the univariate Hawkes process describing mid-point changes of the E-mini S\&P500 and for the trade arrivals of the EuroStoxx in different years, reproduced respectively from Refs.~\cite{Bacry:2014ab} and~\cite{Hardiman:2013aa}. Strikingly, considering the different markets, traded contract, type of events considered, and periods of time considered in these studies, the shape of $\knl[t]$ is almost identical.}
    \label{fig:powlawfit}
\end{figure}

In Ref.~\cite{Bacry:2011kx}, Bacry \etal\ recently introduced a non-parametric 
estimation method for multivariate symmetric Hawkes processes based on 
the spectral factorization of the covariance matrix by the means of the Hilbert 
transform. By calibrating a 1-dimensional Hawkes 
model to the occurrence of trades of the 10 years Euro-Bund future front contract over
75 trading days in 2009, they discovered two important empirical facts: (i) the model
is very close to its stability threshold $||\knl|| = 1$ and 
(ii) the empirical kernel $\knl(t)$ is very well described, over a wide range of scales, 
by the  power-law function Eq.~(\ref{eq:knl_pow_law})
\begin{equation}
\label{pl-kernel}
\knl(t) = \frac{\alpha \beta}{(1+\beta t)^{1+\gamma}} 
\end{equation} 
with $\gamma \simeq 0$.
The first observation directly concerns the level of endogeneity and the one of stability of financial 
markets, a problem, as discussed in the next section, 
that has been addressed afterwards by Filimonov and Sornette or
Hardiman \etal~\cite{Filimonov:2012aa,Filimonov:2013aa,Hardiman:2013aa,Hardiman:2014aa}.
The power-law nature of Hawkes kernels with an exponent $\gamma \simeq 0$ 
has been confirmed by studies that followed, notably
by Hardiman \etal\ on mid-price changes of E-mini S\&P500 futures~\cite{Hardiman:2013aa}
or by Bacry and Muzy on trades arrivals of EuroStoxx index futures~\cite{Bacry:2014ab}.
The plots of Fig.~\ref{fig:powlawfit} are directly extracted from these papers: they
represent in log-log scales the estimated Hawkes kernel for the E-mini SP futures mid-price 
change events  and the EuroStoxx market order occurrences.
One can see that the two estimated kernels, corresponding to different data, different
markets and different estimation methods, are strikingly similar. This suggests
some universality of both $\gamma$ and $C = \alpha \beta^{-\gamma}$ parameters 
in the algebraic decay of Eq.~\eqref{pl-kernel}. 

The origin of this power-law behavior and, in particular, of the values of $\alpha$ and $\gamma$ remains an open question.
Previous empirical results motivated the work by Jaisson~\cite{Jaisson:2014aa} (see also Ref.~\cite{Bacry:2014aa} and the section preceding Eq.~\eqref{eq:approx_criticality}), who proved that, within a particular asymptotics (which includes the $\lone$ norm of the kernel converging to 1), a 1-dimensional kernel with a power-law decay with an exponent $1+\gamma$ (with $\gamma \in ]0,0.5[$) leads to an auto-covariance function of the flow which is power-law with an exponent $1-2\gamma$, i.e., to a long-range dependence of the flow (see also the discussion in Sec.~\ref{ssec:properties}, Eq.~\eqref{eq:asymp_pow_law}). The algebraic decay of Hawkes kernels can then be related to the volatility clustering properties.
Notice that within trading time models~\cite{BouchaudFarmerLillo}, the long-range 
nature of offer and demand has been mostly explained in terms of order splitting dynamics, and is supported by empirical results such as~\cite{Toth2015218}, in which broker-resolved data is analyzed in order to distinguish among the herding and splitting components of the order flow.
It is thus likely that the observed slowly decreasing nature of the kernel directly results from 
the splitting of the meta-orders. Such an hypothesis remains however to be supported by quantitative arguments
and to be confirmed by empirical observations.
Beyond these fundamental questions, the power-law nature of Hawkes kernels remains 
a solid empirical fact which, at least, calls to question all the approaches based 
on exponential Hawkes models.
 
In the same lines of the previously cited studies, Da Fonseca and Zaatour~\cite{Da-Fonseca:2014aa}, perform a parametric estimation (using a GMM approach, see Sec.~\ref{ssec:param-estimation}) of a 1-dimensional Hawkes process with exponential kernels of the form $\alpha\beta \exp(-\beta t)$ on (unsigned) market-order flow data. As expected, market-order clustering translates in a rather low value for $\beta \in [0.02,0.1]$ (depending on the financial time-series) and the $L^1$ norm of the kernel  is found to be very close to criticality, i.e., $\alpha \ge 0.9$ (and very often $\ge 0.95$).
One can also cite the work of Lallouache and Challet~\cite{Lallouache:2014aa}, who performed a maximum likelihood estimation on market orders using a sum of two exponential functions as the Hawkes kernel. They study forex EBS data which are throttled (market orders are gathered within slices of 0.1 seconds) which calls for a rather complex denoising preprocessing. Using goodness of fit tests, they conclude that, though the model performs well when applied to a 1 hour specific intraday time (averaged every day over 3 months), the  intraday seasonality of (mainly) the exogenous intensity $\mu$ does not allow to fit well a whole day.

Hawkes processes have also been used to model extreme price moves at a rather low frequency.
In Ref.~\cite{Embrechts:2011aa}, Embrechts \etal\ study an equally weighted portfolio of 3 indices (Dow Jones, Nasdaq and SP) on an hourly time-frame (on 14 years). Only extreme quantiles of returns are kept (the smallest and the largest $1\%$  quantiles) leading to a 1-dimensional point process whose jumps correspond to an extreme return of any of the three indices. The jumps are then marked using a 3-dimensional marks coding the excess of each index respect to the corresponding $1\%$ quantile. The so-obtained point process is modeled using a 3-dimensional Hawkes process (with an exponential kernel) marked by 3-dimensional Gamma-distributed i.i.d. random variables.
A maximum likelihood estimation is performed and goodness of tests show that this model is well suited for modeling extreme price moves.
Let us point out that, in the same work, a very similar experiment is performed on daily log-returns of an (home-made) index of stocks using this time a 2-dimensional Hawkes process (for coding positive or negative extreme jumps) with 1-dimensional marks.
In the same spirit,
in Ref.~\cite{Chavez-Demoulin:2012aa}, Chavez-Demoulin and Mc Gill constructed a model
for the excesses of an asset price above a given threshold. This model combines a Hawkes process for the excess occurrences with Pareto distributed marks to the excess sizes.
Within this approach, the author's goal was to describe the clustering of large drawdown events through the self-excited dynamics of the Hawkes process. By performing backtests on equities intraday data,  they have shown that this model captures very well extreme
intraday events, notably as compared to standard non-parametric methods
based on extreme value theory.

\subsection{Measuring the endogeneity of stock markets} 
%Models for mid-point price changes such as~\cite{Filimonov:2012aa,Filimonov:2013aa,Hardiman:2013aa}
In a recent series of papers~\cite{Filimonov:2012aa,Filimonov:2013aa,Hardiman:2013aa,Hardiman:2014aa},
some authors addressed, within the framework of Hawkes models, the important problem of the so-called
``{\em volatility puzzle}'', namely the fact that the observed market volatility cannot be explained by classical
economic theory. Indeed, it is well known that prices move too much compared to the flow of pertinent
information that may impact the market. 
This observation naturally leads to the idea that price dynamics is highly endogenous, i.e. mainly driven
by some internal feedback mechanisms. Filimonov and Sornette~\cite{Filimonov:2012aa} were the first to propose a 
quantitative measure of the level of ``market reflexivity''. For that purpose, they model the 
high frequency mid-price variations of some stock index (namely the E-mini S\&P500) as a 1-dimensional
Hawkes process. As explained in Sec.~\ref{ssec:causality},  $||\knl||$ can be interpreted 
as a branching ratio, i.e., the number of events generated by any parent event. Each exogenous event
occurring at rate $\mu$ thus generates $||\knl||/(1-||\knl||)$ events and therefore 
the ratio of endogenous event rate to the overall rate $\Lambda$ in one dimension is, according to Eq.~\eqref{eq:stat_lam},
$$ 
\frac{\mu}{\Lambda} \left(\frac{||\knl|| }{1-||\knl||}\right) = ||\knl|| \; .
$$
This means that $||\knl||$ provides a direct measure of the fraction of endogenous events within the whole population
of mid-price changes and thus a measure of the market reflexivity.
By analyzing the E-mini S\&P500 future contracts over the period 1998-2010, 
Filimonov and Sornette found that the degree of reflexivity has strikingly increased during the last decade.
They suggested that this effect could be directly caused 
by the increasing amount of high frequency and algorithmic trading,
raising the question of the impact of high frequency trading on the market stability.
This analysis has been revisited by Hardiman, Bercot and Bouchaud 
\cite{Hardiman:2013aa} who noticed that Filimonov and Sornette estimation relying on an exponential
parametrization is biased because, as discussed previously, empirical evidences suggest that
Hawkes kernels have a slow (power-law) decay.
Accounting for this feature on their estimation of $||\knl||$, Hardiman \etal\ found that the reflexivity
of the E-mini S\&P future hasn't been increasing during the last decade, but has remained constant at a value
very close to the critical one $||\knl|| = 1$. It is noteworthy that Hardiman \etal\ also provided, in their
study, empirical evidences of the existence of a high frequency cut-off of the Hawkes kernel 
(namely the parameter $\beta^{-1}$ in Eq.~(\ref{pl-kernel}))
that decreases exponentially fast in time and that can be associated with the increase of the trading 
frequency. In a more recent paper, Filimonov and Sornette~\cite{Filimonov:2013aa}, have reviewed all the pitfalls
associated with the estimation $||\knl||$ in the case of a slowly decreasing kernel. They have shown that 
significant biases can be induced by the presence of outliers, edge effects or non stationary effects.
Because some important issues are also related to the way one parametrizes the model (notably
the choice of the high-frequency regularization), Hardiman and Bouchaud~\cite{Hardiman:2014aa} 
proposed a simple non-parametric approximation of the branching ratio $||\knl||$ that relies
on Eq.~(\ref{eq:stat_corr}). Indeed, by considering this equation in $z=0$, it is possible 
to relate the integral of the correlation function, $\lcor[0]$ to $||\knl||$. The number of events $\cnt[T]$ in a window of size $T$, becomes, for $T$ large enough,
$\lcor[0] \simeq T^{-1}\var{\cnt[T]}$
and therefore one gets
\begin{equation}
    \label{eq:branching_est}
    ||\knl|| \simeq 1- \left(\frac{\var{\cnt[T]}}{\av{\cnt[T]}}\right)^{1/2} \; .
\end{equation}
This formula leads to a very intuitive interpretation of the degree of reflexivity: The occurrence of correlated events implies an increase of the variance of $\cnt[T]$ 
with respect to its mean value (for a Poisson process, both quantities are equal so that
one directly gets $||\knl||=0$). 
Using this model free estimator, Hardiman and Bouchaud~\cite{Hardiman:2014aa} 
have confirmed their former claims that the S\&P 500 future appears to have, during the last ten years, 
a stable level of reflexivity, close to the criticality.
Let us notice that this formula only holds for 1-dimensional 
Hawkes processes and has no simple extension in the multivariate situation.

Beyond the debate on the most suitable estimator for the reflexivity parameter and
its genuine behavior, the pioneering work of Filimonov and Sornette  
provided a quantitative
framework allowing to study the endogeneity of market fluctuations with Hawkes processes. 
They notably have shown that such approach can be used
to study particular events such as the flash crashes of April and May 2010~\cite{Filimonov:2012aa}. Their results
may be helpful to devise warning tools in order to anticipate extreme drawdowns
which are of endogenous origin.
The prospects and applications along this path are numerous.
One important question concerns the extension of such studies
by accounting other types of events like e.g. order book events
(see Sec.~\ref{sec:ob_models}).

\section{Price models}
\label{sec:price}
\begin{figure}[htb]
    \begin{center}
        \includegraphics{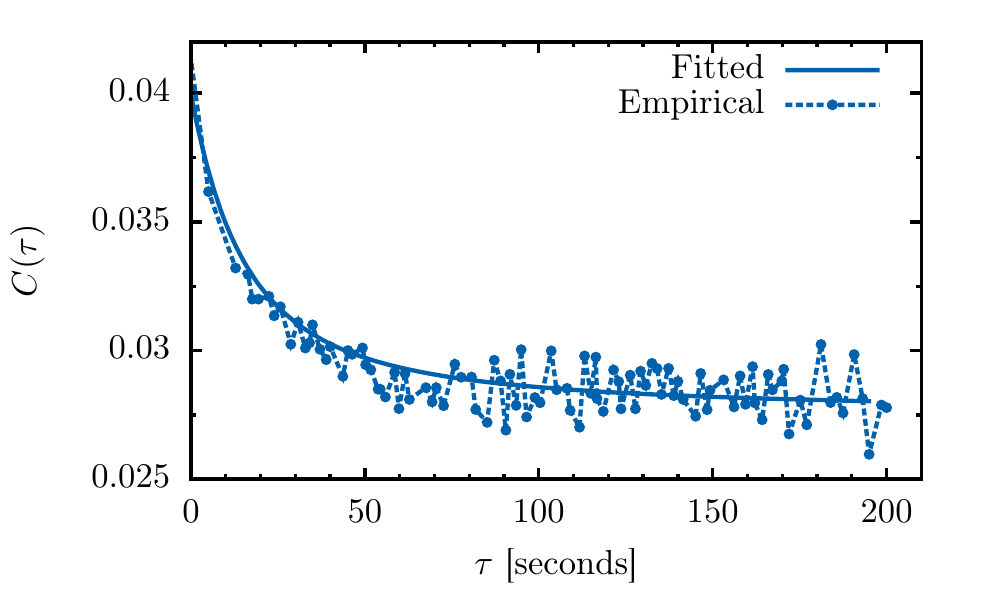}
    \end{center}
    \caption{Reproducing the mid-price behavior at the microstructural level using Bacry \etal\ 2-dimensional Hawkes model. The plots
    	represent the Euro-Bund empirical signature plot as compared to its fit within the model of Bacry \etal. The 
        figure has been reproduced from Ref.~\cite{Bacry:2010}}
    \label{fig:signature}
\end{figure}

Describing the fluctuations of price at the finest time scales, with notably the goal 
of improving volatility and covariance estimations, is a central issue of financial econometrics.
The notion of microstructure noise was considered by many 
authors as an additional noise, superimposed to the standard diffusion, that accounts
for the small scale behavior of the signature plot. Indeed, it is well know that the signature plot
$$
  C(\tau) = \frac{1}{T} \sum_{i=0}^{T/\tau} \left[ P_{(i+1)\tau} - P_{i \tau} \right]^2
$$
that corresponds to the quadratic variation of the mid-price $P_t$ at scale $\tau$, 
strongly increases when $\tau \to 0$ (see Fig.~\ref{fig:signature}). Along the same line, the so-called Epps
effect accounts for the vanishing covariation among the returns of pairs of assets when the return scale $\tau$ goes to zero.
Bacry \etal\  in~\cite{Bacry:2010} proposed as an alternative to standard ``latent price'' models, 
to directly account for the discrete nature of price variations.
They have been the first ones to describe the tick-by-tick variation of the mid-price, $P_t$,  within the framework of Hawkes processes.
In order to do so, these authors considered the two counting processes $N^1_t$ and $N^2_t$ associated
with respectively the arrival times of upward and downward price changes and set:
$$
 P_t = P_0+ N^1_t - N^2_t,~~~t>0
$$ 
where the couple $(N^1_t,N^2_t)$ is a 2-dimensional Hawkes model.
Since, in a first approximation, the dynamics of the upward moves and of the downward moves are expected to be the same,
it is natural to consider a matrix kernel $\Phi$ as the one considered in Eq.~\eqref{eq:kernel_bivariate}
with equal diagonal terms $\knl[(s)][t]$ and anti-diagonal terms $\knl[(c)][t]$. 
It is well known that, at the microstructure level, the price is essentially mean reverting, thus in~\cite{Bacry:2010}, the authors
considered the ``purely mean-reverting scenario'', i.e., the case where $\knl[(s)][t] = 0$ and chose for $\knl[(c)][t]$ an exponential shape.
Within this simple framework, Bacry \etal\ provided a closed-form expression for the signature plot.
Calibrating the model using MLE or GMM estimations on Euro-Bund and Euro-Bobl future data, they have shown
that the model is able to reproduce the scale behavior of the signature plot (see Fig.~\ref{fig:signature}).
Bacry \etal\ also considered a natural extension of the previous model to a 4-dimensional Hawkes model in order to describe the joint
mid-price dynamics of a pair of assets and to reproduce the Epps effect~\cite{Bacry:2010}.
Let us recall that, in Ref.~\cite{Bacry:2013aa}, the authors established that under general conditions, the empirical covariation of a multivariate
Hawkes process converges towards its expected value, that can be easily expressed in terms 
of the Hawkes covariance matrix $\cor[t]$ as given by Eq.~(\ref{eq:stat_corr}). This result allows one, within the Hawkes 
price model of Bacry \etal, to provide analytical expressions for the signature plot, lead-lag behavior and the Epps effect in terms
of Hawkes matrix kernel $\knl$ ~\cite{Bacry:2010,Bacry:2013aa}. 
Let us also mention that, along the line the of previous model, 
in~\cite{Da-Fonseca:2014aa}, the symmetric case was considered, i.e., the ``purely trend following scenario'': 
$\knl[(s)][t] = 0$ and $\knl[(c)][t]$ exponential shape.
The authors compared this scenario with the ``purely mean-reverting scenario'' when used for daily volatility 
estimation using diffusive formula \eqref{eq:ex1diffusive}. 
They showed that the mean-reverting (resp.\ trend-following) scenario underestimates (resp.\ over-estimates) the volatility and concluded 
that a ``full'' model with both cross and self terms is more realistic and should lead to better volatility estimation.
Let us notice that the existence of non-negligible diagonal and anti-diagonal terms has been confirmed by non-parametric estimations 
performed thereafter in~\cite{Bacry:2011kx,Bacry:2014aa,Bacry:2015}.

One also expects the involved kernels not to be exponential functions. Indeed, as shown in many works, (e.g., in Ref.~\cite{Bacry:2011kx}) and as discussed previously in Sec.~\ref{ssec:risk}, empirical self-exciting kernels are closer to a power-law than to an exponential. It is with that consideration in mind that Jaisson and Rosenbaum~\cite{Jaisson:2013aa} developed the diffusive framework that we have already described in second part of Sec.~\ref{ssec:scaling-limit}. It is an alternative framework to the more classical Brownian motion diffusive framework developed in the first part of the same Section. In the second part of their paper, Jaisson and Rosenbaum used their framework to build a version of the 2-dimensional Hawkes model initially introduced by~\cite{Bacry:2010} that converges at large scales towards the Heston price model~\cite{Heston:1993} which displays volatility clustering.
This work can be seen as the very first step towards a ``across scales'' unified model that would fit both microstructure stylized facts of price (i.e., point process with strong mean reversion) and ``diffusive'' stylized facts (volatility clustering and multifractality).
In that sense this is a very promising work.

We can also mention a very interesting generalization of the model introduced by Bacry \etal\ in~\cite{Bacry:2010}. In~\cite{Zheng:2014}, Zheng \etal\ introduced a model for the coupled dynamics of the best bid and ask prices. 
It starts by coding each best price using the Bacry \etal\ model leading to a 4-dimensional price model. The main difficulty comes from the fact that one needs to encode in the model the fact that the ask price needs to lie strictly above the bid price. This is achieved through a spread point process whose dynamics is coupled with the dynamics of both best prices. It is used to measure the distance (in ticks) between the ask price and the bid price: the spread is increased (resp.\ decreased) by 1 each time either the ask (resp.\ bid) component jumps upward (resp.\ downward) or the bid (resp.\ ask) component jumps downward (resp.\ upward). Finally a non-linear term is introduced in the Hawkes model: the intensities of the downward (resp.\ upward) jumps of the ask (resp.\ bid) price are set to 0 as soon as the spread process is equal to 1.
The authors of~\cite{Zheng:2014} developed a whole new rigorous framework for this constrained, non-linear Hawkes model in which they were 
able to establish several properties (including a diffusive limits). They perform maximum likelihood estimation on real data (using exponential kernels) and show that they were able to reproduce rather well the signature plot.

Let us finally cite the work of Fauth and Tudor in \cite{Fauth:2012} where the authors proposed to describe bid and ask prices
of an asset (or a couple of assets) within the framework of marked multivariate Hawkes model. Motivated by the empirical 
observation (performed on high-frequency Euro/USD and Euro/GPB FX rates from 30-01-2012 to 10-03-2012) that, as the transaction
volumes increase, the inter-trade durations decrease, the authors proposed to consider, in addition to an exponential Hawkes kernel, a multiplicative mark (the function $\chi$ in Sec.~\ref{ssec:marked}) 
that corresponds to a power-law function of the volumes: $\chi(v) = C v^{\nu}$. 
Their model thus describes the events corresponding to an increase/decrease of 
the bid/ask as a four dimensional Hawkes process marked by transaction volumes. 
By adding suitable constraints in order to avoid infinite spread, they calibrated
the model on FX rates data by a maximum likelihood approach. Fauth and Tudor have shown
that their model is consistent with empirical data by reproducing  the signature plots of the considered assets and 
the behavior of the high-frequency pair correlation function (Epps effect).

\section{Impact models}
\label{sec:impact}
\subsection{Market impact modeling}
Market impact modeling is a longstanding problem in market microstructure literature and is
obviously of great interest both for theoreticians and  practitioners (see e.g.,~\cite{BouchaudImpact} for a recent review). While for the former market impact reflects the mechanism enforcing the efficiency of markets, allowing prices to reflect fundamental information, for the latter it represents a cost which needs to be carefully minimized when executing an order. For a trader, market impact induces extra costs per transaction which needs to be added to the fees charged directly by the market, forcing him to split large orders and trade them incrementally in sequences of smaller child orders. Any of such sequences of orders is called a \emph{meta-order}, and quantifying their effect on prices is at the heart of market-microstructure regulation discussions.

The theory of market price formation and the relationship between the order flow and price changes has made significant progress during the last decade thanks to the increasing availability of intraday data~\cite{BouchaudFarmerLillo}. Many empirical studies have provided evidence that the
price impact has, to many respects, some universal properties and is the main source of price variations. This corroborates the picture of an ``endogenous'' nature of price fluctuations that contrasts with the classical scenario according to which an ``exogenous'' flow of information drives the prices towards a 
fondamental value~\cite{BouchaudFarmerLillo}.

\begin{figure}[ht!]
\begin{center}
        \includegraphics{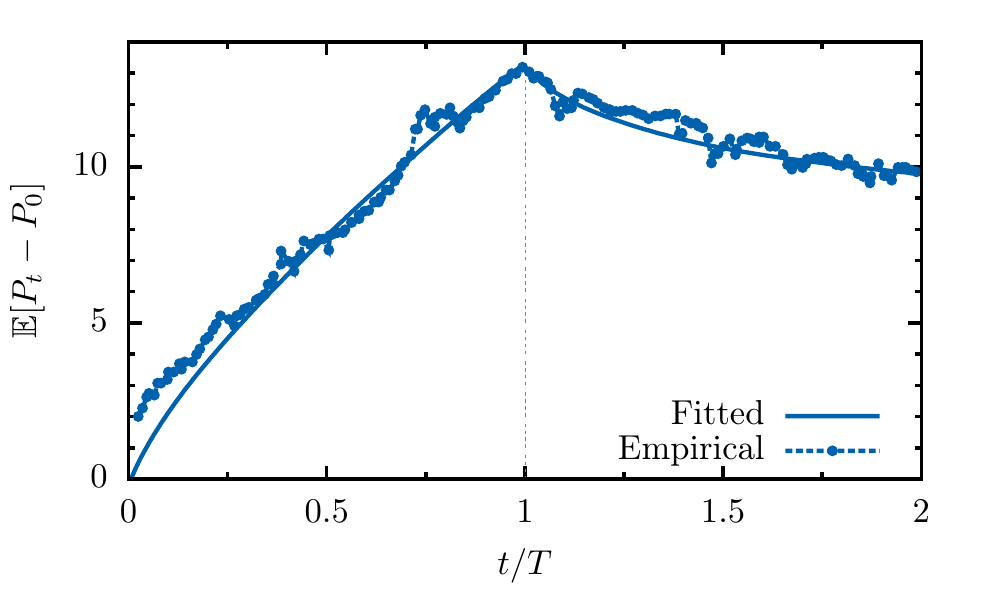}
\end{center}
    \caption{% 
    Averaged empirical market impact curves (normalized in time) over a large database of broker meta-orders.
    The fit is performed using the impulsive HIM model. This figure has been reproduced from~\cite{Bacry:2014c}.
    \label{fig:impact}
    }
\end{figure}

If a meta-order is placed at time $t=0$ and executed until time $t=T$, an associated market impact curve can be defined by a proxy of the price variation it directly or indirectly causes\footnote{We focus here on the impact of meta-orders, rather than with the impact of individual orders, or with the one of trade imbalance, which have also received considerable interest in the literature.}. 
One generally distinguishes two phases: an increasing (concave)
part during the execution of the meta-order (i.e., on the interval $[0,T]$), followed by
a decaying (generally convex) resilient part. The existence of permanent impact, i.e., a non zero
asymptotic value (for large time) of the market impact curve, is a central problem that remains under debate.
The typical shape of a market impact curve is shown in Fig.~\ref{fig:impact}, that was was obtained by Bacry \etal\ in~\cite{Bacry:2014c} by averaging empirical impact curves
over a large database of broker meta-orders.
Let us point out that this type of measure of market impact cannot be obtained using anonymous market data
(i.e, one cannot easily identify the meta-order of a given agent). This, together with the extremely slow intensity of the price signal with respect to its statistical fluctuations, is the reason why empirical results in this respect have been obtained only in relatively recent years.

Bacry and Muzy proposed recently a price impact model based on Hawkes processes~\cite{Bacry:2014aa}. 
They suggested to directly account for the joint dynamics of mid-price 
and the market order occurrences. More precisely the four dimensions of their Hawkes 
model correspond to the mid-price upward and downward jumps and the buying and selling
market order flow (they did not account for the volume of the orders nor the price jump sizes).
The matrix kernel $\knl$ can then be decomposed into four sub-blocks of size $2 \times 2$. 
The first one describes the self-excitement of the market
order flow, the second one the self-excitement of the price, the third one the (market) impact of the trades 
on the price while the last one accounts for the feedback influence of price moves on market
order flow intensity. By calibrating the model directly
from anonymous high-frequency data (using the Wiener-Hopf non-parametric estimation technique described in App.~\ref{ssec:nonparam} on the most liquid
maturity of EuroStoxx and Euro-Bund future contracts over 800 
trading days from 2009 to 2012), the authors of~\cite{Bacry:2014aa} were able 
to disentangle the self and cross excitation dynamics of mid-price changes
from the impact of market orders.
In particular, they have shown that 
the market 
impact block is mainly diagonal: buying (resp.\ selling) orders are mostly triggering
upward (resp.\ downward) price moves. Moreover the shape of the diagonal impact function
is very localized around $t=0$. This means that a market order no longer directly impacts 
the price after a very short delay (i.e. less that 0.1$s$). 
They have also shown that the feedback sub-block is mostly anti-diagonal with slightly negative diagonal kernel functions (see Sec.~\ref{ssec:non_linear} for a short discussion on negative kernels). That indicates that an upward (resp.\ downward) jump in the price tends to increase the intensity of the selling (resp.\ buying) market order flow and to decrease the intensity of the buying (resp.\ selling) market order flow.
The shape of kernels involved in the self-excitation of the market order flow or the price jumps confirms former estimations performed within lower dimensional models: long-range correlation of the signs of the trades and mainly long-range mean-reversion of the price.
All these results were confirmed (using a database with a much higher precision in time) 
in the work of~\cite{Bacry:2015} (see Sec.~\ref{ssec:level1}).
Bacry and Muzy established that, within their framework, it is possible to determine
the entire impact profile of some meta-order that is built by accounting for the ``bare'' direct
localized marked impact and by a ``dressed'' impact that involves mainly the trades power-law self-excitation. 
They provided analytical expressions for this curve and notably established in both the increasing and the  decreasing
phase of the impact function, its relationship with the power-law behavior of the self-exciting kernel of market order events.
Using the previously described empirical findings, they have shown that one can recover the typical 
shape depicted in Fig.~\ref{fig:impact}.

In~\cite{Bacry:2014c}, an impact model based on the 2-dimensional Hawkes price model described in Example \ref{example1} of Sec.~\ref{ssec:defin} (in which only mean-reversion influence has been kept, i.e.,  $\knl[(s)][t] = 0$) has been introduced. It takes into account the impact of an exogenous (buying) meta-order strategy $r(t)$, where $r(t)$ corresponds to the trading rate of the (buying) strategy per unit of time (so $r(t)\dd t$
corresponds to the number of shares bought between time $t$ and $t+\dd t$). More precisely, the so-obtained Hawkes Impact Model (HIM) writes
\begin{equation}
\label{eq:impact}
\lambda^1_t= \mu +  \convknl[(s)] \star \dd\cnt[2][t] +  \convknl[(I)] \star f(r_{t})~~{\mbox{and}}~~ \lambda^2_t= \mu + \convknl[(s)] \star \dd\cnt[1][t] + \convknl[(x)] \star f(r_{1}),
\end{equation}
where $\dd\cnt[1][t]$ (resp.\ $\dd\cnt[1][t]$) codes the upward (resp.\ downward) jumps of the price and
 $f(r(t))dt$ (with $f(0) = 0$) codes the infinitesimal impact of a buy order of volume $r(t)dt$. The function $f$ corresponds to  the {\em instantaneous} impact function and $\convknl[(I)]$ and $\convknl[(x)]$ correspond respectively to the impact kernel and the cross-impact kernel (this latter describes the impact of a buying order on downward jumps, of course, we expect that $||\convknl[(x)]|| << 
||\convknl[(I)]||$)\footnote{As for $(s)$ and $(c)$, $(I)$ stands for \emph{impact} and $(x)$ for \emph{cross-impact}.}. Following the empirical findings of ~\cite{Bacry:2014aa}, the impulsive-HIM model corresponds to the particular choice of an impulsive (very localized) impact kernel 
$\knl[(I)][t] = \delta(t)$, i.e.,  a Dirac distribution. 
 Moreover, as for the choice of $\convknl[(x)]$, the impulsive-HIM model considers that the market reacts to the newly arrived order as if it triggered an upward jump: 
$ \knl[(x)][t] = {C} \frac{\knl[(s)][t]}{||\convknl[(s)]||}$.
The constant $C>0$ is a very intuitive parameter that quantifies 
the ratio of ``contrarian'' reaction (i.e.\ impact decay)
and of the ``herding'' reaction (i.e.\ impact amplification). Analytical formula for the market impact curve were obtained and three cases of interest
 for $C$ were distinguished in~\cite{Bacry:2014c}: $C=0$
corresponds to no contrarian reaction (strong permanent impact), $C=1$ corresponds 
to a  contrarian reaction as ``strong'' (in terms of the norm) as the herding one (no permanent impact) and finally
$C\in]0,1[$  corresponds to a contrarian reaction which is not zero but strictly smaller than the herding reaction. Fig.~\ref{fig:impact} shows a fit of the empirical market impact curve using this model with 
a power-law microstructure kernel ($\knl[(s)][t] \sim t^{-\gamma}$, when $t\rightarrow +\infty$), in which case, the market impact curve is proven to be decaying to the permanent market impact value with a power-law $t^{-(\gamma+1)}$.
 % {\bf ?? What is the relation with other gamma? Is it $\gamma\to\gamma+1$?}.  

Let us point out, that in a totally different framework, Jaisson~\cite{Jaisson:2014aa}, linked, in the asymptotic limit defined by Jaisson and Rosenbaum in~\cite{Jaisson:2013aa}
(see Sec.~\ref{ssec:scaling-limit}), the power-law exponent of the self-excitement kernel involved in the market order flow
and the power-law exponent of the market impact decay.
He derived his results,
within a 2-dimensional Hawkes model (with only self-excitement kernels) for the market order flow,
 from (i) a price martingale hypothesis  and (ii) a linear market impact hypothesis. 

\subsection{Optimal execution}
A natural application of price impact models is to 
define optimal liquidation strategies.
Hewlett \cite{Hewlett:2006aa} was the first to address this problem using Hawkes models.
He proposed to model the occurrence of buy and sell market orders on FX markets using a bivariate exponential Hawkes process. He found that these events are mostly self-excited, the cross excitation intensity between buy and sell events being negligible. Using Eq.~\eqref{eq:pred_implicit}, Hewlett determined
the expected future trade imbalance that, within a linear price impact model, 
allows one to determine the expected future price returns and the associated risk.
He then showed that this approach allows one to 
devise a liquidation strategy that maximizes a mean-variance utility function.

Application of Hawkes model for optimal execution has also been considered more recently
by Alfonsi and Blanc~\cite{Alfonsi:2014aa} who modelled the price process
using a linear impact of liquidity takers. 
More precisely, the price is decomposed a the sum of a 
``fundamental'' price which variations are 
proportional to a fraction of the trade imbalance and 
a ``transient'' price that is moved by the remaining fraction
of the trade imbalance but with a damping term that represents 
the market resiliency caused by the market makers behavior.
Alfonsi and Blanc have provided explicit expressions of optimal liquidation strategy
(i.e.\ the one with the minimum expected cost)
when the flow of buy/sell market orders that impact the price is either
a Poisson process or a 2-dimensional Hawkes process with a symmetric 
exponential kernel matrix. They notably show that price manipulation
strategies (i.e. liquidation strategies with negative expected cost)
always exist for a Poisson model while they can be excluded in the Hawkes
model provided its parameters meet some specific conditions.
According the these conditions, the self-excitation should exactly compensate 
the price resiliency so that resulting price is a martingale. The other condition
leads to identify the fraction of endogenous orders within the Hawkes model 
with the proportion of market orders involved in the transient part of the 
price behavior.

\section{Orderbook models}
\label{sec:ob_models}
Modeling faithfully the occurrence of various type of orders in the order book with
the aim of understanding the mechanisms at the origin of price formation,
volatility and liquidity variations is probably the main challenge that
the applications Hawkes processes in financial econometrics have to face.
Although this goal is far from being reached, 
some authors have already tackled the problem and made significant 
progress on these issues.

\subsection{Level-I book models}
\label{ssec:level1}
The Level-I book description concerns the events that 
exclusively occur at the best bid and best ask levels of the order book.
Even if this approach discards most of the book information, since best bid or
best ask values are directly related to the asset mid-price and market
orders mostly impact the book at level-I, one can expect that this level
of description is rich enough to capture most of the market features. The study of Biais \etal~\cite{Biais:1995aa} confirms indeed that most of the activity of an order book takes place close to the best quotes, while Cont \etal~\cite{Cont:2013ab} indicate that a substantial part of the dynamics of prices can be accounted for by the evolution best bid and best ask only.

One of the first applications of Hawkes models to order-book modeling at level-I was performed
by Large~\cite{Large:2007aa} who formalized the concept of book resiliency, namely
the ability of the order book to replenish after being depleted by a large trade.
Large suggested to quantify resiliency by the way large trades alter future intensities
of order occurrences. For that purpose he introduced the response kernel $G^{ij}(t)$:
\begin{equation}
\label{eq:defrespk}
G^{ij}(t) \dd t =
\avc{ \dd\cnt[i][t]}{\filt{0},\dd\cnt[j][0]=1}-
\avc{ \dd\cnt[i][t]}{\filt{0}} \,.
\end{equation}
$G^{ij}(t)$ simply describes the increase of the future conditional intensity
of events of type $i$ caused, directly or indirectly, by the the occurrence, at time $t=0$ of an event of type $j$.
In the case of a Hawkes process of kernel matrix $\Phi$, 
Large proved that $G^{ij}$ satisfies the integral equation
in Eq.~(\ref{eq:inv_kernel}) defining $\ikn(t)$.
In other words, the matrix $\ikn(t)$ can be interpreted as the increase 
of the expected number of events after a lag $t$ caused
by the occurrence of some event at time $0$.
Large then considered order book data from LSE, modeled as a 10-dimensional
Hawkes processes where the book events are classified according to whether they move or not the mid-price:
market and limit orders that move the mid-prices (4 components if one distinguishes bid and ask),
market and limit orders that leave the book unchanged (4 components) and cancel orders (2 components). 
The impact of ``aggressive'' orders  on the rate of forthcoming
events is then studied. The processed data consisted 
in LSE stock data (Barclays equity) timestamped at the resolution of 1$s$ during the 22 trading days 
of January 2002.
Large used MLE estimation within the class of exponential kernels in order to estimate $\ikn$.
His results allowed him to provide a ``causal'' interpretation (Large prefers the term 
``precipitation'' than ``cause'') of the main event occurrence in the book. He mainly found
that aggressive limit orders are principally caused by aggressive market orders, measuring thereby
the market resiliency in all his aspects, magnitude, trade direction and characteristic time. 
Consistently with former studies, Large estimated that the studied stock value is resilient less than
40 \% of cases and, when it is the case, the book replenishment occurs within a time frame 
of around 20$s$. He also shown that market order dynamics is mostly self-excited
and correlated over a large time. Aggressive market orders are also triggered by aggressive limit
orders as a consequence of the ``race to liquidity''.

\begin{figure}[htb]
    \begin{center}
        \includegraphics{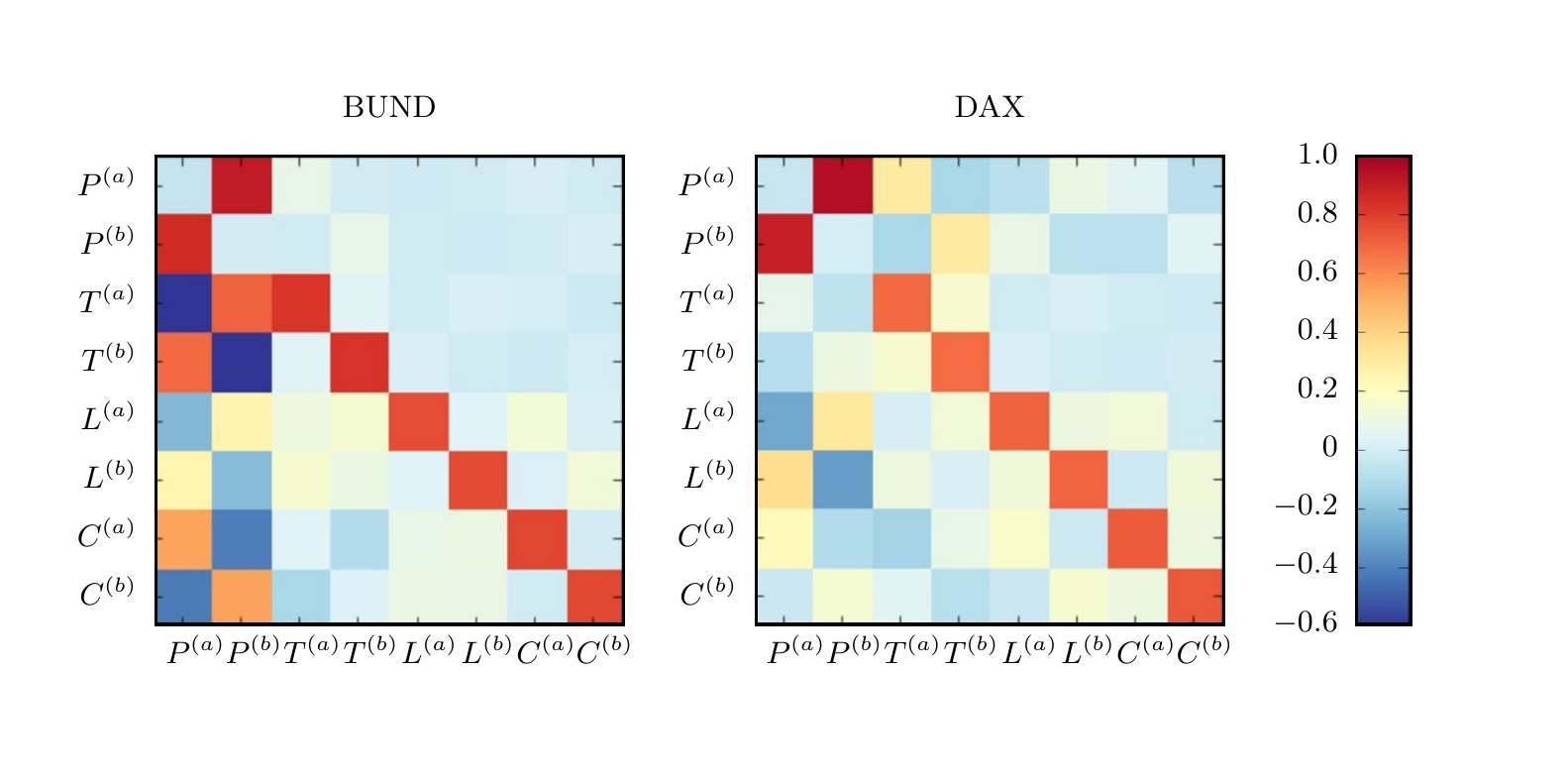}
    \end{center}
    \caption{Empirical determination of the matrix of Hawkes kernel norms in the 8-dimensional model of level-I book events.
        $P$ stands for mid-price change events, $T$ for trade events, $L$ for limit order events and $C$ for cancel events. The superscripts $(a)$ 
        $(b)$ indicates the direction, ask or bid of the events. This figure is reproduced from~\cite{Bacry:2015}.}
    \label{fig:phimatrix}
\end{figure}

A similar analysis of level-I order book data was recently conducted by
Bacry \etal~\cite{Bacry:2015}. These authors made a slightly different categorization than
Large and distinguished all book events (market, limit and cancel orders) 
that leave the mid-price unchanged (6 components accounting for bid and ask sides) 
from events that move the mid-price up or down (2 components). The dynamics
of these event occurrence has then been modeled as a 
8-dimensional Hawkes process. Unlike Large, Bacry \etal\ performed a non-parametric estimation 
of the matrix of kernels using the method described in Sec.~\ref{ssec:nonparam}. They considered book data
time-stamped at a time resolution of $10^{-6} s$ and the analysis has been performed
up to lags of a few minutes. The event dynamics has thus been considered over a range of time scales close to 
eight decades. The main result reported by the authors is that the book event dynamics is mainly
self-exciting, except for the mid-price changes for which cross-excitation effects are strongly 
dominating. This is illustrated in Fig.~\ref{fig:phimatrix} where the values of the estimated kernel norms are reported using a color map: one can see that the resulting matrix is mainly diagonal except
in the price sub-block, that is anti-diagonal. 
The observation that price changes events are mainly triggered by price change events
is in agreement of Large previously reported results. Moreover, Bacry \etal\ observations 
confirmed the previously reported empirical fact that this triggering effect is mostly 
anti-diagonal, i.e., present price changes impact future price changes in the opposite direction.
Previous findings using low-dimensional model concerning the kernel shapes were also confirmed:  
the market is highly endogenous, whatever the type of event one considers and all the dominating kernels
(the diagonal kernels for orders leaving the price unchanged and the anti-diagonal kernels for mid-price moves)
are slowly decreasing, well described by a power-law behavior as in Eq.~\eqref{pl-kernel}.
The richness of the Hawkes model of~\cite{Bacry:2015} allowed the authors to account for a richer dynamical behavior than previous works and to describe and quantify the high-frequency influences
between all types of events. They notably characterized the impact of price changes
on the book event flow, a quantity that turns out to be very sensitive to the asset tick size
(as estimated by the probability that the mid-price has to move). They also provided evidence of some inhibitory
effects which result from negative values of some Hawkes kernels. For example it was observed that,
for a large tick asset (like the Euro-Bund futures), an upward price move not only triggers forthcoming trades at bid but also inhibits trades on the ask side.  

Restricting previous Large model to market order flows at best bid and best ask,  Muni-Toke and Pomponio 
\cite{Muni-Toke:2012aa} studied the dynamics of trade-through orders. A market order is a trade-through if part of it is executed at the next best limit. Thus, for that purpose, they used the 2-dimensional Hawkes process (with exponential kernels) described by Eq.~\eqref{eq:ex1diffusive}.
Though very rare for some assets (e.g., Euro-Bund future), trade-through can be quite frequent on other contracts. For instance on the BNP stock ~\cite{Muni-Toke:2012aa} finds an average of 400 trade-throughs per day.
Parametric estimation using MLE on exponential kernels have been performed on Euronext stocks restricting intraday-time to 9h30 to 11h30 am to avoid very strong intraday seasonality effects. Goodness of fit tests confirm that, as for the full market order flow, the main components of the 2-dimensional Hawkes kernel are the self-exciting kernel.

\subsection{Full order book model}
\label{ssec:full_ob_model}
In~\cite{Toke:2010}, Muni Toke generalized the zero-intelligence model introduced in~\cite{farmer2003doubleauction} to the Hawkes framework.
While in the latter model the authors built a full model for the order book in which all the involved flows (i.e., limit and market orders at any level) are independent pure Poisson processes. It is clearly a very rough approximation, market orders are known to be long-range dependent (see Sec.~\ref{ssec:risk}) due to the splitting of large meta-orders.
Moreover, one expect limit orders to be also highly auto-correlated as well as correlated with the market-order flow due to the fact market makers interact with market takers. 
In~\cite{Toke:2010}, a 2-agents model is introduced. It comprises the following agents:
\begin{itemize}
    \item A \emph{liquidity provider}: 
    \begin{itemize}
        \item Arrivals of limit orders are modeled by using either a pure homogeneous Poisson process or a 1-dimensional Hawkes process with an exponential kernel (cancellations are treated as in the zero-intelligence model, i.e., each order has a life-time which is an i.i.d. exponential random variable using whose parameter is fixed a priori)
        \item The price-levels of new limit orders are randomly chosen by first choosing the (bid or ask) side with probability 1/2 and then by sampling their values from a Student distribution
    \end{itemize}
    \item A \emph{liquidity taker}: 
    \begin{itemize}
        \item Modeled using either a pure homogeneous Poisson process or a 1-dimensional Hawkes process with an exponential kernel.
    \end{itemize}
\end{itemize}
All the volumes of both limit and market orders are i.i.d.\ and exponentially distributed variables. Hawkes processes use exponential kernels which are estimated parametrically using Maximum Likelihood Estimation.

\begin{figure}[p]
    \begin{center}
        \includegraphics{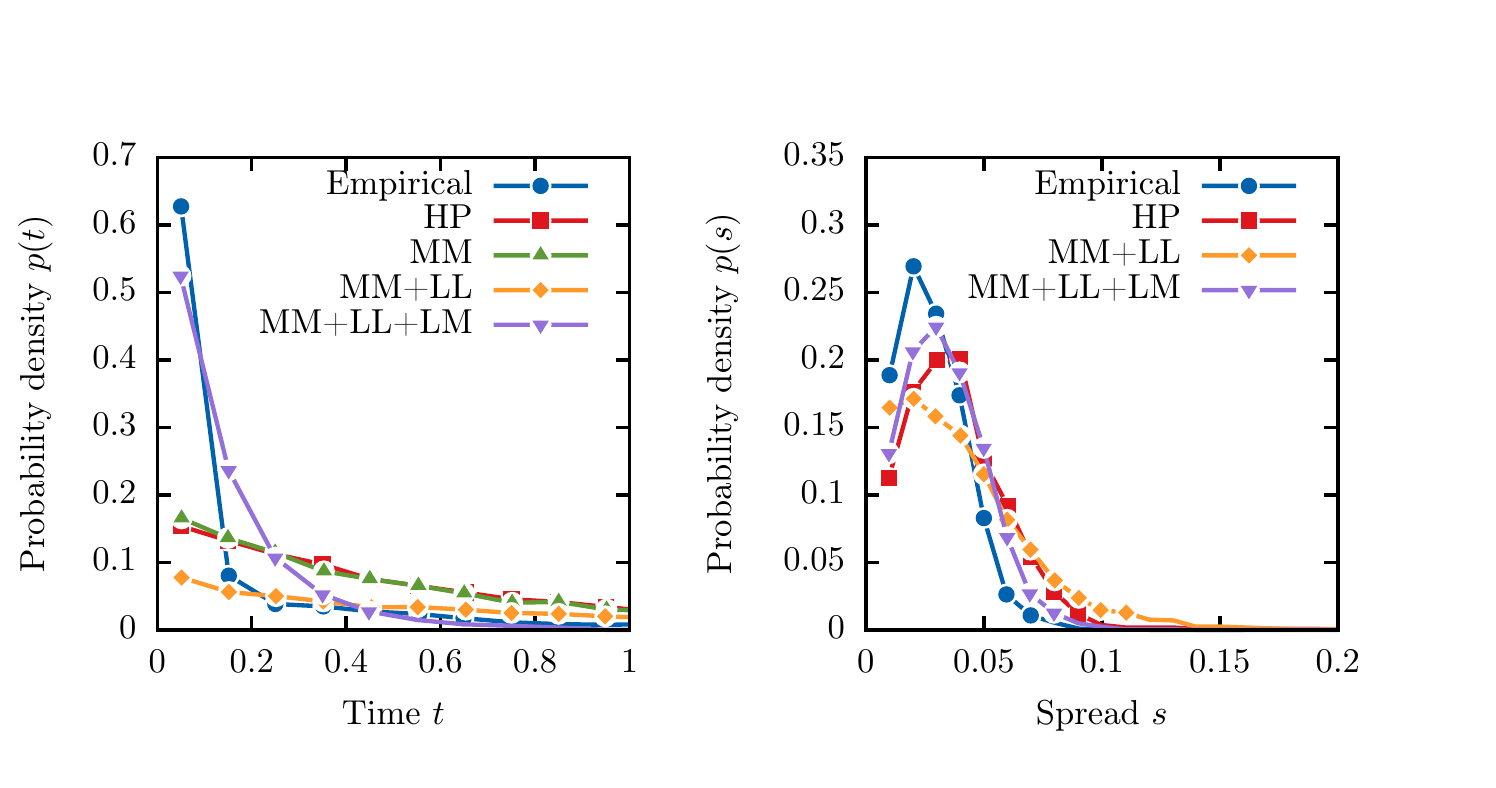}
    \end{center}
    \caption{
        (Left panel) Empirical density function of the distribution of the duration between a given market order and the first arrival of a limit order
        after that market order. The density function is displayed for empirical BNPP time-series, and for three different models that were fitted (using maximum likelihood) on these data, namely: a purely Homogeneous Poisson model (HP) for both market and limit orders, a model with a pure Poisson process for limit orders  and a 1-dimensional Hawkes process for the market orders (MM), a model with two independent 1-dimensional Hawkes processes for limit and market orders (LL+MM) and finally this last model with a cross-exciting term which characterize the influence of past market orders on future limit orders (LM).
        (Right panel) Empirical density function of the distribution of the bid-ask spread. The density function is displayed for empirical BNPP time-series, and for three different models that were fitted (using maximum likelihood) on these data, namely: a purely Homogeneous Poisson model (HP) for both market and limit orders, a model with a pure Poisson process for limit orders  and a 1-dimensional Hawkes process for the market orders (MM), a model with two independent 1-dimensional Hawkes processes for limit and market orders (LL+MM) and finally this last model with a cross-exciting term which characterize the influence of past market orders on future limit orders (LM).  These figures are reproduced from Ref.~\cite{Toke:2010}.
        }
    \label{fig:toke}
\end{figure} 

Not surprisingly, inter-trade times of market orders are shown to be much better modeled by a 1-dimensional Hawkes process (referred to as MM since the only Hawkes kernel involved is one which deals with the influence of Market orders on themselves) than by a pure Homogeneous Poisson process (HP). 
In the same way, inter-time of limit orders are shown to be much better modeled by a 1-dimensional Hawkes process (LL) than by HP.
Along the same lines, the left panel of Fig.~\ref{fig:toke} shows that inter-time between a given market order and the first arrival time of a limit order after that market order is much better reproduced by a 2-dimensional Hawkes (referred to as MM+LL+LM) model with self-exciting kernels for both market (MM) and limit orders (LL)  and a single cross-exciting kernel which corresponds to the influence of past market orders on future limit orders (LM). The figure shows that a model with two independent Hawkes models (MM+LL) with no cross-exciting kernels (LM) does not perform well. 
In~\cite{Toke:2010}, Muni-Toke claims that the other cross-exciting kernel (the one describing the influence of past limit orders on future market orders) is negligible. 
Thus the order book dynamics seems to be driven mainly by the liquidity taker agent rather than the liquidity provider, i.e.,  in a first approximation, the market makers strategy consists basically in  reacting to liquidity takers, whereas liquidity takers take decisions independently from market makers. 

Finally the right panel of Fig.~\ref{fig:toke} shows the so-obtained distribution of the bid-ask spread for empirical data, the pure Homogeneous Poisson (HP) model, the MM+LL Hawkes model and the MM+LL+LM model.
Again, this latter model is the one which best fits the empirical data.

Let us mention a theoretical work by Jedidi and Abergel \cite{jedidi:hal-00821607} in which a Hawkes-based markovian framework for the whole order-book dynamics is studied.  

\section{Other models}
\label{sec:other_models}

\subsection{Systemic risk models}
\label{ssec:systemic_risk}
Hawkes models have been also be used at coarser time scales where asset prices
are mainly diffusive. They can be used in mixed models in order, for instance, to account
for jumps that occur over the diffusion process.
This is the spirit of the model developed by A\"{\i}t-Sahalia \etal~\cite{Sahalia:2010hx} that proposed to describe the contagion of a crisis across all the world markets by superimposing a multivariate self-excited Hawkes process to a standard multivariate continuous diffusion model.
According to this model, called be the authors, ``Mutually exciting Jump-Diffusion'', the log-price
vector satisfies:
\begin{equation}
    \label{eq:syst_risk}
    \dd X_t = \mu_t + \sigma_t \dd\wien[t] + Z_t \dd \cnt[t]
\end{equation}
where $\wien[t]$ is a $\nn$-dimensional Brownian motion, $\sigma_t$ is a stochastic volatility 
and $\cnt[t]$ is a $\nn$-variate Hawkes process
that accounts for the self-excited nature of price jumps occurrence  ($Z_t$
is a random variable accounting for the direction and the intensity of the jump).
A GMM estimation procedure is proposed in~\cite{Sahalia:2010hx} based on a closed-form
expressions of some moments associated with the returns variations in
the univariate and bivariate cases with exponential Hawkes kernels.
This estimation has been applied to five international equity indices data associated with 
respectively US, Europe, Asian, Pacian and Latin America zones.
The authors found that the jumps terms have significant self-excitation 
components and as far as the ``contagion'' effect is concerned,
it seems that the US equities are the ones with the greatest
influence on other markets.

In Ref.~\cite{Errais:2010aa}, Errais \etal~proposed to model credit default events
in a portfolio of securities as correlated point processes. More specifically they 
considered that the dynamics of such events is described by a marked Hawkes process
with exponential kernels. A stated by Prop.~\ref{prop:markov}, 
in this case, the couple of processes
$(\lam,\cnt)$ is a Markov process. Thanks to the Dynkin formula, the authors
provided explicit expressions for the conditional distribution
of both the marked and counting processes. This result 
was used to price portfolios if credit derivatives such as index and tranch swaps. 
Their model was then calibrated from index and swap data during September 2008 
that witnessed several credit default
notably Lehman-Brothers default. The authors have shown that by,
capturing the dependence of default events, their model provided good fits of market 
data, unlike standard approaches that failed during this period.
Errais \etal~emphasized that marked Hawkes models with exponential kernels 
belong to the more general class of affine point processes introduced by 
Duffie \etal~\cite{Duffie:2003}. An affine point process involves a stochastic intensity vector
which is a Markov process with drift, diffusion and jump
terms. Errais \etal~\cite{Errais:2010aa} have shown that their approach remains tractable within 
the class  of affine point processes that provides a richer jump interaction structure.

Let us also quote the recent similar work of Dassios and Zhao \cite{Dassios:2011} who
addressed the question of default risk modeling and contagion propagation within the
framework of the so-called ``Dynamic Contagion Process'' that is an exponential marked 
Hawkes model but where the Poisson exogenous events (the ``immigrants'') are replaced by a shot-noise (constructed
as the first generation of the same exponential Hawkes model with a different mark probability density).
The authors have established the theoretical distributional properties of this new process (that remains a Markov process)
and provided analytic expressions for the its probability generating function. As in Errais \etal~\cite{Errais:2010aa},
they showed that their approach is particularly suitable 
for modeling the dependence structure of arriving events with dynamic contagion
and proposed an application to credit risk,

\subsection{Accounting for news}
\label{ssec:news}
In~\cite{Rambaldi:2014aa}, Rambaldi \etal\  used Hawkes processes for modeling the impact of news on the EBS (foreign exchange market) quotation time-series, consisting in a list of quotation timestamps. Let us note that, since EBS data are throttled (aggregated below a window of 0.1$s$), a  randomization procedure has been used for homogenization.
The model for the quotation arrival times is a 1-dimensional Hawkes model involving either a double exponential or a power-law (expressed as a sum of 15 exponentials so to have a convenient Maximum Likelihood Estimation procedure). 
The impact of the news (on the quotation time-series) can be seen 
as particular instances of localized non stationarities. Thus, Rambaldi \etal\ introduced an 
exogenous kernel $\phi^{(news)}$ (an exponential function) that accounts for the impact of a particular instance of a news, leading to the model~\footnote{Formally, this model can be seen as a 1-dimensional version of the 2-dimensional price impact model of Eq.~\eqref{eq:impact} in which $f(r_t)$ is replaced by the Dirac distribution centered at time $t_0$, $\delta(t-t_0)$.}: 
\begin{equation}
\lambda_t = \mu + \phi\star \dd\cnt[t] + \phi^{(n)}(t-t_0),
\end{equation}
where $t_0$ is the time of occurrence of a particular news.
Let us point out that $\phi^{(n)}$ (where $(n)$ stands for \emph{news}) is allowed to have a non-causal component (i.e., $\phi^{(n)}(t)$ is non zero for $t<0$) in order to  account for anticipation effects. 
Estimation is performed (using regular Maximum Likelihood Estimation procedure) on a 3 hour period around the considered news. Though the results are very noisy, the authors showed that the model captures nicely both the amplitude and the time scale of the news effect.
The distribution of the  $\lone$ norm $||\phi^{(n)}||$ for the different news has broad distribution (that goes beyond 1) that clearly reflects the diverse effects of news on the market. Moreover, using some proxies for quantifying how unexpected a particular news is, Rambaldi \etal\ 
clearly showed that the norm $||\phi^{(n)}||$ is not only related to the news impact but also to its degree of surprise.

\subsection{High-dimensional models}
\label{ssec:high_dim}
\paragraph{Modeling co-jumps.}
In~\cite{Bormetti:2013}, Bormetti \etal~modeled the complex dynamics of the extreme returns in a basket of stocks.
They started by elaborating a rather elaborate procedure for identifying extreme returns (referred to as ``jumps'' in the paper, corresponding to anomalous values) from 1-minute returns time-series. On a basket of $N=20$ Italian stocks they found up to 280 jumps on a single stock on the overall period (of 88 days) and up to a total of 505 jumps per day on all the stocks.
They clearly established that some jumps arrive at the same time, i.e., within the same time-window. This prevents from modeling the overall process as an $N$-dimensional Hawkes process: Within such a framework, co-jumps cannot occur without introducing an impulsive component inside the kernel $\knl$.
Bormetti \etal\ finally suggested the following model 
\begin{itemize}
\item A 1-dimensional Hawkes process is used for modeling the co-jumps arrival times. The kernel of the Hawkes process is chosen to be a sum of exponential functions.
\item Each time $t$ a co-jump occurs, each stock $i$, independently one form each other,  has a probability $p_i$ to jump. These probabilities are estimated empirically from real data. 
\item For each stock $i$, a 1-dimensional Hawkes process is used to model the idiosyncratic jumps of that stock, i.e., the jumps that are do not correspond to co-jumps. Again, the kernels of the Hawkes process is chosen to be a sum of exponential functions.
\end{itemize}
In their paper, Bormetti \etal\ developed a precise procedure to perform estimation of this model which is shown to be quite robust. Moreover they  showed that it is able to capture simultaneously the time clustering of jumps and the high synchronization of jumps across assets.

\paragraph{Clustering with graph models.}
In~\cite{Scott:2014}, Linderman and Adams developed a probabilistic model that combines Hawkes processes with random graphs models, that they applied on S\&P100 data. 
Each component of the Hawkes kernel codes the changes (of more than $0.1\%$) in the last traded price 
of a given asset during a whole week. Thus a 100-dimensional point process with 182.037 events is obtained. The kernels are chosen 
as
\begin{equation}
\knl[ij][t] = A^{ij} W^{ij} h_{\theta_{ij}}(t),
\end{equation}
where $A$ is a random binary (0 or 1) valued matrix, $W$ a random matrix with positive entries and $h_{\theta}(t)$ a parametric kernel (of parameters $\theta$) such that $\int h_\theta = 1$  (a logistic normal density with two parameters is chosen). 
Intraday seasonality is modeled using an exogenous intensity of the form
$\mu = m + \nu e^{y(t)}$ where $\nu$ is a constant matrix and $y(t)$ is univariate Gaussian process.
The random graph model is used to reflect the probability of the different network structures through the prior distributions of the matrices $A$ and $W$. They basically depend on a latent distance which is chosen in $\R^2$ imposing an overall sparsity ($20\%$) and a characteristic distance scale. Thus each stock $k$ corresponds to a latent coordinate $x^k$ in  $\R^2$ that is estimated through a fully-Bayesian, parallel inference algorithm.
A figure is displayed where each stock is represented as a dot in the latent 2-dimensional space with a color coding corresponding to its corresponding sector (among six sectors). Linderman and Adams showed that
some sectors, notably energy and financial, tend to cluster together, indicating an increased probability of interaction between stocks in the same sector. Other sectors, such as consumer goods, are broadly distributed, suggesting that these stocks are less influenced by others in their sector.
One can think that this approach will pave the way, with the framework of multivariate Hawkes processes, to very promising
applications where one will process large amounts of data associated with a great number of interacting components/agents in order
to shed new light on the complexity of financial markets.

Finally, let us note that Mastromatteo and Marsili~\cite{mastromatteo2011criticality} tried to overcome the problem of the estimation of a high-dimensional Hawkes process by mapping it onto a graphical (Ising) model, so to describe the clustering of trading times in a set stocks. They reconstructed the interaction network of the $\nn=100$ most traded stocks of the NYSE during the year 2003 by using machine learning techniques, and reported an overall scaling of the weight matrix $W \propto \nn^{-1}$. Their study evidences the presence of a large collective market mode of the matrix $A$, driving the overall level of market activity close to the critical point.

\section{Concluding remarks}
Hawkes processes are extremely versatile processes that can be considered as the building blocks 
for modeling the occurrence of time-correlated discrete events, playing the same role that Auto-Regressive
models have for describing continuous-valued signals. Within a relatively simple mathematical framework,
they allow one to characterize precisely the interactions between different categories of events
and to account for their causal relations. From the statistical point of view, they can be generated
with relatively simple simulation algorithms, while several efficient estimation methods exist to calibrate them. 

Hawkes processes have already proven to be  very useful in many domains like, e.g., the modeling
of earthquakes, neuronal or social network activity.
Because they allow one to describe data at the resolution of individual events and to account 
for endogenous triggering, contagion and cross-excitation phenomena, 
they have naturally found applications in the field of high-frequency finance.
In this paper, we have proposed an overview of many of such applications that concern
a wide variety of problems. From the question of the level of endogeneity of financial markets to the modeling of order-book events, including impact, risk contagion modeling, the design of optimal execution strategies,
we saw that Hawkes models have been used to address a large spectrum of issues.

Far from being exhaustive, this review is necessarily a ``snapshot'' of a topic that 
is promised to a rapid evolution and growth. Many of the studies we mentioned can be considered
as pioneering works that bode more important results and a deeper understanding 
of the market dynamics at the microstructural level.
Among the promising routes that will be
explored in forthcoming studies, some can be easily anticipated.
Since several market stylized facts at medium or large time scales
seem to originate from the market microstructure,  
the micro-to-macro transition is an important issue. In that respect, 
understanding the long-time behavior of the Hawkes process and the instabilities that can emerge
can be of great importance. One may hope not only to recover known models at larger times, but also to account for genuinely new phenomena like microstructural crashes. 
Beyond the Brownian limit which one expects in usual
situations, elucidating the emerging properties of the Hawkes process close to the instability limit seems a promising area of research due to the richness of the model in the vicinity of the critical point and also because empirical data 
seem to indicate that markets operate close to this instability threshold.

The study and the estimation of Hawkes model in the high-dimensional regime is also a challenging prospect since the complexity of financial markets 
results notably from the interaction between a very large 
number of components (market participants, agents, assets, information fluxes,...).
There have been many progresses made recently in that direction notably under
the impetus of the community of big-data and the studies of viral diffusion across social networks.
One can expect many interesting applications of theses approaches to high-frequency finance to be proposed in the next future.

\begin{appendices}

\section{Table of financial applications found within each discussed paper}
\label{sec:table}
Each of the academic works discussed throughout our paper involving numerical experiments on financial data is listed in the following table. 
Apart from the column names that are explicit: 
$D$ stands for dimension of the Hawkes model, $T$ for the length of the historical data used to calibrate the model, $\dd t$ for the time resolution of the data, $\Phi(t)$ for the shape of the Hawkes kernel. $N$-Exp stands for a sum of $N$ exponentials, PL for ``power-law '' and NP for ``non-parametric''.
% \newgeometry{left=1cm,right=1cm}
% \thispagestyle{empty}
% \begin{table}[h]
% \input{pubs.tex}
% \end{table}
% \restoregeometry

%%% CHANGED :
% \restoregeometry conflicted with \linewidth of successive \begin{itemize} environment,
% which then required to be changed to \begin{itemize}[rightmargin=54.75012pt]

\begin{wtable}[h]
\thisfloatpagestyle{empty}
\input{pubs.tex}
\end{wtable}

\clearpage

\section{Simulation}
\label{sec:simul}

There are two main frameworks for simulating Hawkes processes: an intensity-based framework and a cluster-based framework, depending on whether it makes direct use of the intensity regression Eq.~\eqref{eq:hawkes} or of the cluster representation of the Hawkes process (see Sec.~\ref{ssec:cluster_repr}). The most popular method being certainly the intensity-based thinning method initially introduced by 
Lewis~\cite{Lewis:1979} in the general context of non homogeneous Poisson processes, and modified later by Ogata~\cite{Ogata1981}.
\paragraph{Thinning.} The thinning algorithm is an incremental procedure in which the successive jumping times are generated sequentially.
It can be easily adapted to the multi-dimensional context and can be generalized to the marked case.
In its simplest version (valid for decreasing kernels), given a starting time $t$ it basically consists in
sampling a candidate jumping time $t+\Delta t$ using an exponentially distributed random variable $\Delta t$
with parameter
$\lam[(tot)][t] = \sum_{k=1}^D \lam[k][t]$.
Then a random variable $U$ is uniformly drawn in the interval $[0,\lam[(tot)][t]]$.
If $U<\lam[(tot)][t]-\lam[(tot)][t+\Delta t]$,
the jump is  rejected. If not rejected,
the jump is assigned
to the component $i$, where $i$ is the largest index satisfying $U\ge \lambda_t-\sum_{k=1}^{i-1} \lam[k][t+\Delta t]$. Let us point out that the complexity of the algorithm can be substantially reduced in the case of exponential kernels.

\paragraph{Time-change} Another common intensity-based approach for simulating a non-homogeneous Poisson process $N_t$ uses the following well known fact: if one defines the cumulative intensity function $F_t = \int_0^t \lambda_u du$, then $N_{F^{-1}_t}$ is an homogeneous Poisson process of intensity 1.
Thus in order to perform simulation one needs to know how to simulate $F^{-1}_{t_{m+1}}-F^{-1}_{t_{m}}$ where $t_{m+1}-t_{m}$ is exponentially distributed. This algorithm has been applied for instance in the case of exponential kernels in~\cite{Dassios}. In fact, in this simpler case it is possible to invert analytically the function $F$.

\paragraph{Cluster algorithm} The cluster approach consists basically in simulating the branching structure described in Sec.~\ref{ssec:cluster_repr}. Hence, the algorithm is sequential in the generations index $n$ rather then in real time $t$. One can see for instance~\cite{Moller} which describes how one can perform simulation of marked Hawkes processes using the branching structure. 
A short survey and a more comprehensive list of references can be found in the recent work~\cite{Dassios}.

\section{Statistical inference}
\label{sec:estimation}
\subsection{Parametric estimation}
\label{ssec:param-estimation}

\paragraph{Maximum Likelihood Estimation.} The most commonly used technique for parametric estimation of Hawkes processes is the Maximum Likelihood Estimator (MLE), which has been first introduced by Ogata in~\cite{ogata_linear_1982}.
The log-likelihood of a non-homogeneous, multi-dimensional Poisson defined as in Eq.~(\ref{eq:hawkes}) reads
\begin{equation}
\label{eq:ll}
\log \mathcal{L}(\exo,\knl ) = - \sum_{i=1}^D \int_0^T \dd t \, \lambda_t^i + \sum_{m=1}^M \log \lam[k_m][t_m] ,\
\end{equation}
where the couples $\{(t_m,k_m)\}_{m=1}^M$ denote respectively event times and components.
Given a Hawkes kernel $\knl_\theta$ depending upon a set of parameters $\theta$, it is possible to estimate it from Eq.~(\ref{eq:ll}) by solving the problem
\begin{equation}
    (\mu^*,\theta^*) =  \operatornamewithlimits{\mbox{argmax}}_{(\mu,\theta)}\log \mathcal{L}(\exo,\knl_\theta ) \,.
\end{equation}
In the case of a general Hawkes process, the computation of the likelihood (or its gradient) is of the order of $O(M^2D)$ where $M$ is the total number of jumps. The fact that it reduces to $O(MD)$ when the kernels are exponential functions is one of the major reasons why exponential kernels (or sum of exponential kernels) are so commonly used in a parametric framework~\cite{Embrechts:2011aa,Toke:2010,Muni-Toke:2012aa,Bormetti:2013,Rambaldi:2014aa,Zheng:2014}. 

\paragraph{EM based Estimation.}
In~\cite{Veen:2008}, the authors use an Expectation Maximization (EM) based technique, an iterative procedure comprising the alternation of two steps of \emph{expectation} (E) and \emph{maximization} (M), in order to exploit the cluster representation of the Hawkes process described in Sec.~\ref{ssec:cluster_repr}. Given a parametrization of the Hawkes kernel $\knl_\theta$, and a set of parameters $(\mu,\theta)$, each iteration consists in:
\begin{itemize}
\item {\bf E-step} $(\mu,\theta)\to \{ p_{m m'}\}$: Estimating, for all ordered pair of jumps $(t_m,t_{m'})$ (with $t_m < t_{m'}$),
 the probability $p_{m m'}$ that, within the cluster framework described in Sec.~\ref{ssec:cluster_repr}, $t_m$ is an ancestor of $t_{m'}$.
 \item {\bf M-step} $\{ p_{m m'}\} \to (\mu,\theta)$: Computing the model parameters $(\mu,\theta)$ from the probabilities $\{ p_{m m'}\}$ estimated in the E-step. 
 \end{itemize}
This algorithm converges very quickly when the product of the average intensity $\Lam$ and the characteristic timescale $\tau$ of the support of $\Phi$ is small, i.e., $\Lambda \tau \ll 1$, meaning that few jumps are detected in intervals of size $\tau$.

Let us point out that General Method of Moments (GMM) can be also used~\cite{Da-Fonseca:2014aa}, based, for instance, on auto-covariance analytical formula such as \eqref{eq:stat_cor_exp_explicit}. 

\paragraph{High-dimensional estimation.} All the previous techniques cannot be applied in a large dimensional context (e.g., $D \ge 100$) without substantial modifications. 
For instance, even in the case of ``simple'' exponential kernels, the number of parameters is of the order of $D^2$, so that applications to contexts in which $D$ is large become unfeasible due to overfitting and/or computational issues. In order to address the issue of parametric estimation in large dimensions, one needs to use algorithms which involve regularization. Within the last year, Hawkes processes in large dimension have been the subject of numerous academical papers (see for instance~\cite{Zhou}). Most of them, adapt more or less ``classical'' convex optimization techniques in the 
framework of Hawkes processes with exponential kernels of the form $\alpha^{ij} \beta^{ij}e^{-\beta^{ij} t}$. Let us point out that, in all these papers, in order to get a convex log-likelihood \eqref{eq:ll}, the parameters $\beta^{ij}$ are a priori fixed (i.e., they are not estimated), and generally not chosen to depend on $i$ or $j$. Thus, in that case, the kernel matrix $\Phi$ can be written in matrix form as $\Phi = \alpha \beta e^{-\beta t}$, where $\alpha$ is the so-called (weighted) adjacency matrix $\alpha = \{\alpha^{ij}\}_{1\le i,j\le D}$. $\alpha$ describes the ``connections'' between the different components of the Hawkes process. In order to solve the so-obtained convex-optimization of the log-likelihood, penalizations terms on $\alpha$ are customarily introduced.
They are generally of two sorts: an $L^1$ penalization that ensures sparsity and a trace-norm 
penalization that ensures low-rank. To the best of our knowledge, these types of algorithms have only been used once in the context of finance (see Sec.~\ref{ssec:high_dim} about~\cite{Scott:2014}).
Clearly, one can easily forecast that there will be, in the near future, a huge number of new algorithms for parametric estimation in large dimensions, some of them, improving the state of the art.

\subsection{Non-parametric estimation}
\label{ssec:nonparam}
Very few non-parametric estimators of the kernel matrix of a Hawkes process have appeared in the academic literature so far. We review them in the next section.

\paragraph{EM based estimation.}
Historically, the first one~\cite{Lewis:2011aa}, corresponds to a non-parametric version of the EM estimation algorithm described in the previous section. It is based on regularization (via $L^2$ penalization) of the method initially introduced by~\cite{Marsan:2010aa} in the framework of ETAS model for seismology.
It has been developed for 1-dimensional Hawkes processes. 
The maximum likelihood estimator is computed using the same two steps as in the parametric case. The E-step corresponds to estimating all the $p_{m m'}$ probabilities and the M-step corresponds to estimating $\phi(t)$ and $\mu$ from these probabilities.
In~\cite{Lewis:2011aa}, some numerical experiments on particular cases are performed successfully even when the exogenous intensity
$\mu$ depends slowly on time: the whole function $\mu(t)$ is estimated along with the kernel $\phi(t)$ with a very good approximation.
However, as explained in~\cite{Bacry:2014ab}, this method has two main drawbacks:
\begin{itemize}
\item The convergence speed of the EM algorithm drastically decreases when the decay speed of the kernel $\phi(t)$ is low (e.g., power-law decaying kernel),
\item The probabilistic interpretation of the kernel values involved in the EM method prevents the kernels to have negative values (see Sec.~\ref{ssec:non_linear}).

% \footnote{From a mathematical point of view, negative kernels can be considered in the framework of non-linear Hawkes processes \eqref{eq:non_lin_hawkes} where the function $h(x) =x$ for positive $x$ and 0 elsewhere.}
\end{itemize}

\paragraph{Contrast function based estimation.}
In a recent series of papers~\cite{Rey0,Rey1}, some authors proposed, within a rigorous statistical
framework, a second approach for non-parametric estimation. It  relies on the minimization of the
so-called $L^2$ {\em contrast function}.
Given a realization of a Hawkes process $\tilde N_t$ on an interval $[0,T]$, associated with the parameters $(\tilde \mu,\tilde \Phi(t)$), the estimation is based on minimizing the {\em contrast function} $C(\mu,\Phi)$:
\begin{equation}
\label{contrastmin}
   (\mu^*,\Phi^*) =  \operatornamewithlimits{\mbox{argmin}}_{(\mu,\Phi)} C(\mu,\Phi),
\end{equation}
where
\begin{equation}
\label{contrast}
   C(\mu,\Phi) =   \left( \sum_{i=1}^D \int_0^T \lambda_t^i(\mu,\Phi)^2 \dd t -  2\int_0^T \lambda_t^i(\mu,\Phi) \dd\tilde N^i_t  \right)
\end{equation}
and
\begin{equation}
   \lambda_t^i(\mu,\Phi) = \mu^{i} + \sum_{j=1}^D\int_{-\infty}^t \Phi^{ij}(t-s) \dd\tilde N_s^j \; .
\end{equation}
Let us point out that, minimizing the expectancy of the contrast function is equivalent to minimizing the $L^2$ error on the intensity process. Indeed, if ${\cal F}_t$ is the information available up to time $t^-$, since $\avc{\dd\tilde N_t^i }{\filt{t}} = \tilde \lambda_t^i \dd t$, one has
\begin{eqnarray}
\label{eq:l2}
\operatornamewithlimits{\mbox{argmin}}_{(\mu,\Phi)}
\av{C(\mu,\Phi)} & = & \operatornamewithlimits{\mbox{argmin}}_{(\mu,\Phi)} \sum_{i=1}^D \left( \av{\lambda^i(\mu,\Phi)^2}-
\av{\lambda^i(\mu,\Phi) \tilde \lambda^i}\right) \\
& = & \operatornamewithlimits{\mbox{argmin}}_{(\mu,\Phi)} \sum_{i=1}^D \av{(\lambda^i(\mu,\Phi)-\tilde {\lambda^i})^2}.
\end{eqnarray}
The minimum value (zero) is of course uniquely reached for $\Phi = \tilde \Phi$ and $\mu =\tilde \mu$.

In~\cite{Rey1}, the authors chose to decompose $\Phi$ on a finite dimensional-space
(in practice, the space of the constant piece-wise functions)
 and to solve directly the minimization problem \eqref{contrastmin} in that space. For that purpose, in order to regularize the solution (they are essentially working with some applications in mind
for which only a
small amount of data is available, and for which the kernels are known to be well localized), they chose to penalize the minimization  with a Lasso term (which is well known to induce sparsity in the kernels), i.e., the $L^1$ norm of the components of
$\Phi$.
Let us point out, that minimizing the contrast function and minimizing the expectancy of the contrast function are two different stories. The contrast function is stochastic, and nothing guarantees that the associated linear equation is not ill-conditioned.
In~\cite{Rey1}, the authors prove that, under certain conditions on $\Phi$, the linear equation is invertible, i.e., the associated random Gram matrix is almost surely positive definite.
 In practice (they study real signals from neurobiology), they choose the components of $\Phi$ to be piece-wise constant.

\paragraph{The Wiener-Hopf approach.}
Let us point out that, since $\lambda^i_t$ is expressed linearly in terms of  $\mu^i$ and of the $\{\knl[ij][t]\}_{ij=1}^\nn$, minimizing the $L^2$ error~\eqref{eq:l2} is equivalent to solving
a linear equation, which is nothing but the Wiener Hopf Eq.~\eqref{eq:wiener-hopf}.
Consequently, minimizing the expectancy of the contrast function is equivalent to solving the Wiener-Hopf equation.
In~\cite{Bacry:2014aa,Bacry:2014ab}, the authors chose to perform non-parametric estimation by directly inverting this system.
They have proved that, as long as $g(t)$
 corresponds to a conditional intensity of a Hawkes process,  \eqref{eq:wiener-hopf} has a unique solution in $\chi(t)$ which is the kernel matrix $\Phi(t)$. The algorithm uses quadrature technique in order to discretize the system.
It can be summarized in the following way: 
 \begin{itemize}
\item Estimation of the vector $\Lambda$ (simply using as an estimator of $\Lambda^i$ the number of jumps of the realization of $N^i$ divided by the overall time realization)
\item Non-parametric estimation of the matrix $g(t)$ defined by \eqref{eq:cond_int} (using kernel density estimation techniques)
\item Fix the number of quadrature points (e.g., gaussian quadrature) to be used as well as the support for all the kernel functions. 
\item Discretize the Wiener-Hopf system on the quadrature points and inverse it.
 This leads to an estimation of the kernel functions on the quadrature points. Estimation on a finer grid as well as $\lone$ norm can be obtained through simple quadrature formula.
 \item Estimation of the exogenous intensity $\mu$ using \eqref{eq:stat_lam}.
\end{itemize}
We refer the reader to~\cite{Bacry:2014ab} where the full algorithm (including the methodology for choosing the bandwidth value for density kernel estimation as well as the number of quadrature points) is described precisely along with the comparisons with the other methods presented above. 
It has actually been improved in~\cite{Bacry:2015} to take particular care of slowly decreasing kernels. 

As compared to the approach developed in~\cite{Rey1},
this algorithm is better adapted to the case in which a large amount of data is available.
As compared to the EM approach, 
we shall advocate its use over the EM algorithm mainly in two cases (which are often satisfied when dealing with financial data):
\begin{itemize}
\item Either the kernel functions  are not localized (e.g., power-law). Indeed, in that case the EM algorithm is known to be very slow to converge (see~\cite{Lewis:2011aa} and~\cite{Bacry:2014ab}),
\item Or some of the kernel functions have negative values (see second point on the EM algorithm above). We refer the reader to~\cite{Rey1} for justification why Wiener-Hopf approach allows to deal with negative kernels. 
 \end{itemize}

Finally, let us point out that, in the particular case the kernel matrix is known to be symmetric (which is always true if the dimension $D=1$), the method developed  in~\cite{Bacry:2011kx} uses a spectral method
for inverting \eqref{eq:stat_corr} and deduces an estimation from the second-order statistics. It can be seen as a particularly elegant way of solving the Wiener-Hopf equation.

\end{appendices}

% {\bf ???
%~\cite{Wheatley:2014aa}
% Estimation of the Hawkes Process With Renewal Immigration Using the EM Algorithm ?????
% WE NEED TO READ THIS PAPER 
% }

\section*{Acknowledgements}
We thank J.P.~Bouchaud and S.~Hardiman for providing data from Refs.~\cite{Hardiman:2013aa,Hardiman:2014aa}.
This research benefited from the support of the ``Chair Markets in Transition'', under the aegis of ``Louis Bachelier Finance and Sustainable Growth'' laboratory, a joint initiative of \'Ecole Polytechnique, Universit\'e d'\'Evry Val d'Essonne and F\'ed\'eration Bancaire Fran\c{c}aise.

\bibliographystyle{plain}
\bibliography{review_hawkes}

\end{document}

%% file: notation.tex
\newcommand{\etal}{{\em et al.}}
\newcommand{\id}{\mathbb{I}}
\newcommand{\R}{\mathbb{R}}
\newcommand{\ind}[1]{\mathds{1}_{#1}}
\newcommand{\dd}{{\rm d}}
\newcommand{\nn}{D}
\newcommand{\av}[1]{ {\mathbb{E}\left[\,{#1}\,\right]} }
\newcommand{\var}[1]{ {\mathbb{V}\left[\,{#1}\,\right]} }
\newcommand{\avc}[2]{ {\mathbb{E}\left[\,{#1}\,\big\vert {#2} \right]} }
\newcommand{\filt}[1]{ {\mathcal{F}_{#1}} }
\newcommand{\lone}{{L^1}}
\NewDocumentCommand\knl{oo}{
    \ensuremath{
    \IfNoValueTF{#2}{
    \Phi\IfNoValueTF{#1}{}{(#1)}
    }{
    \phi^{#1}(#2) }
    }
}
\NewDocumentCommand\convknl{oo}{
    \ensuremath{
    \phi^{#1}
    }
}
\NewDocumentCommand\ikn{oo}{
    \ensuremath{
    \IfNoValueTF{#2}{
    \Psi\IfNoValueTF{#1}{}{(#1)}
    }{
    \psi^{#1}(#2) }
    }
}
\NewDocumentCommand\lkn{oo}{
    \ensuremath{
    \IfNoValueTF{#2}{
    \hat \Phi\IfNoValueTF{#1}{}{(#1)}
    }{
    \hat \phi^{#1}(#2) }
    }
}
\NewDocumentCommand\lik{oo}{
    \ensuremath{
    \IfNoValueTF{#2}{
    \hat \Psi\IfNoValueTF{#1}{}{(#1)}
    }{
    \hat \psi^{#1}(#2) }
    }
}
\NewDocumentCommand\knlmat{o}{
    \ensuremath{
        \phi\IfNoValueTF{#1}{
        }{
        ^{#1}
        }
    }
}
\NewDocumentCommand\lknmat{o}{
    \ensuremath{
        \hat \phi\IfNoValueTF{#1}{
        }{
        ^{#1}
        }
    }
}
\NewDocumentCommand\cnt{oo}{
    \ensuremath{
        N\IfNoValueTF{#2}{
            \IfNoValueTF{#1}{}{_{#1}}
        }{
        ^{#1}_{#2}
        }
    }
}
\NewDocumentCommand\wien{oo}{
    \ensuremath{
        W\IfNoValueTF{#2}{
            \IfNoValueTF{#1}{}{_{#1}}
        }{
        ^{#1}_{#2}
        }
    }
}
\NewDocumentCommand\mart{oo}{
    \ensuremath{
        Y\IfNoValueTF{#2}{
            \IfNoValueTF{#1}{}{_{#1}}
        }{
        ^{#1}_{#2}
        }
    }
}
\NewDocumentCommand\lam{oo}{
    \ensuremath{
        \lambda\IfNoValueTF{#2}{
            \IfNoValueTF{#1}{}{_{#1}}
        }{
        ^{#1}_{#2}
        }
    }
}
\NewDocumentCommand\exo{o}{
    \ensuremath{
        \mu\IfNoValueTF{#1}{
        }{
        ^{#1}
        }
    }
}
\NewDocumentCommand\Lam{o}{
    \ensuremath{
        \Lambda\IfNoValueTF{#1}{
        }{
        ^{#1}
        }
    }
}
\NewDocumentCommand\Sig{o}{
    \ensuremath{
        \Sigma\IfNoValueTF{#1}{
        }{
        ^{#1}
        }
    }
}
\NewDocumentCommand\exotd{oo}{
    \ensuremath{
        \mu\IfNoValueTF{#2}{
            \IfNoValueTF{#1}{}{(#1)}
        }{
        ^{#1}(#2)
        }
    }
}

\NewDocumentCommand\cnd{oo}{
    \ensuremath{
        g\IfNoValueTF{#2}{
            \IfNoValueTF{#1}{}{(#1)}
        }{
        ^{#1}(#2)
        }
    }
}

\NewDocumentCommand\alp{o}{
    \ensuremath{
        \alpha\IfNoValueTF{#1}{
        }{
        ^{#1}
        }
    }
}

\NewDocumentCommand\bet{o}{
    \ensuremath{
        \beta\IfNoValueTF{#1}{
        }{
        ^{#1}
        }
    }
}

\NewDocumentCommand\lcnd{oo}{
    \ensuremath{
        \hat g\IfNoValueTF{#2}{
            \IfNoValueTF{#1}{}{(#1)}
        }{
        ^{#1}(#2)
        }
    }
}

\NewDocumentCommand\cor{oo}{
    \ensuremath{
        c\IfNoValueTF{#2}{
            \IfNoValueTF{#1}{}{(#1)}
        }{
        ^{#1}(#2)
        }
    }
}
\NewDocumentCommand\lcor{oo}{
    \ensuremath{
        \hat c\IfNoValueTF{#2}{
            \IfNoValueTF{#1}{}{(#1)}
        }{
        ^{#1}(#2)
        }
    }
}

%% file: pubs.tex
\begin{longtable}{rlrrrrrrr} 
& & \multicolumn{7}{c}{\bf Fitted Hawkes} \\ 
\cline{3-9}& & $D$ & $N_t$ & Contracts & $T$ & $\mathrm{d}t$ & $\Phi(t)$ & Comments \\ 
\endhead
\endfoot
\hline 
\multicolumn{2}{l}{\bf Section 2} &&&&&& \\ 
\cite{Bowsher:2007aa} & Bowsher & 1-2 & Trade & Stocks & 2 months & 1s & 2-Exp & + Seasonality\\ 
\cite{Chavez-Demoulin:2012aa} & Chavez {\em et al.} & 1 & Extreme $\Delta P$ & Stocks & 1 year & 1ms & 2-Exp & Marked\\ 
\cite{Jaisson:2013aa} & Jaisson {\em et al.} & 8 & Level-I & Futures & 1 year & 1$\mu$s & NP & \\ 
\hline 
\multicolumn{2}{l}{\bf Section 3} &&&&&& \\ 
\cite{Bacry:2011kx} & Bacry {\em et al.} & 1 & Trade & Futures & 3.5 months & 1ms & NP & \\ 
 & &2 & Price & Futures & 3.5 months & 1ms & NP & \\ 
\cite{Bacry:2014ab} & Bacry {\em et al.} & 1 & Trade & Futures & 4.5 years & 1ms & NP & \\ 
\cite{Da-Fonseca:2014aa} & Da Fonseca {\em et al.} & 1 & Trade & \begin{tabular}{@{}r@{}}Stocks\\Futures\end{tabular} & 2 years & 1ms & Exp & \\ 
 & &2 & Mid-Price & \begin{tabular}{@{}r@{}}Stocks\\Futures\end{tabular} & 2 years & 1ms & Exp & \\ 
\cite{Embrechts:2011aa} & Embrechts {\em et al.} & 2 & Extreme $\Delta P$ & Indices & 17 years & 1day & Exp & Marked\\ 
 & &1 & Extreme $\Delta P$ & Indices & 14 years & 1h & Exp & 3d-Marked\\ 
\cite{Filimonov:2012aa} & Filimonov {\em et al.} & 1 & Mid-Price & Futures & 13 years & 1s & Exp & \\ 
\cite{Filimonov:2013aa} & Filimonov {\em et al.} & 1 & Mid-Price & Futures & 15 years & 1s & Exp & \\ 
 & &1 & Mid-Price & Futures & 15 years & 1s & PL & \\ 
\cite{Hardiman:2013aa} & Hardiman {\em et al.} & 1 & Mid-Price & Futures & 14 years & 1ms & N-Exp & \\ 
 & &1 & Mid-Price & Futures & 1 year & 1ms & NP & \\ 
\cite{Hardiman:2014aa} & Hardiman {\em et al.} & 1 & Mid-Price & Futures & 16 years & 1s & Exp & \\ 
 & &1 & Mid-Price & Futures & 16 years & 1ms & NP & \\ 
\cite{Lallouache:2014aa} & Lallouache {\em et al.} & 1 & Trade & FX & 3 months & 0.1s & 2-Exp & \\ 
\hline 
\multicolumn{2}{l}{\bf Section 4} &&&&&& \\ 
\cite{Bacry:2010} & Bacry {\em et al.} & 2 & Price & Futures & 2 hours & 1ms & Exp & \\ 
\cite{Fauth:2012} & Fauth {\em et al.} & 4 & Trades & FX-Rates & 2 months & 1ms & Exp & Marked\\ 
\cite{Zheng:2014} & Zheng {\em et al.} & 2 & Trade & FX & 10 days & 0.1s & Exp & Constrained\\ 
\hline 
\multicolumn{2}{l}{\bf Section 5} &&&&&& \\ 
\cite{Bacry:2014aa} & Bacry {\em et al.} & 4 & Price+Trade & Futures & 3.75 years & 1ms & NP & \\ 
\cite{Bacry:2014c} & Bacry {\em et al.} & 2 & Price & Stocks & 1 year & 1ms & PL & \\ 
\cite{Hewlett:2006aa} & Hewlett & 2 & Trades & FX & 2 months & 1s & Exp & \\ 
\hline 
\multicolumn{2}{l}{\bf Section 6} &&&&&& \\ 
\cite{Large:2007aa} & Large & 10 & Level-I & Stocks & 22 days & 1s & Exp & \\ 
\cite{Muni-Toke:2012aa} & Muni Toke {\em et al.} & 2 & Trade-through & Stocks & 5 months & 1ms & Exp & \\ 
\cite{Toke:2010} & Muni Toke & 2 & Orderbook & Stocks & 15 days & 1ms & Exp & \\ 
\hline 
\multicolumn{2}{l}{\bf Section 7} &&&&&& \\ 
\cite{Bormetti:2013} & Bormetti {\em et al.} & $1 \times20$ & Price & Stocks & 88 days & 1 mn & Exp & Co-jump\\ 
\cite{Errais:2010aa} & Errais {\em et al.} & 1 & Default & Credit & 21 days & 1day & Exp & \\ 
\cite{Scott:2014} & Linderman {\em et al.} & 100 & Extreme $\Delta P$ & Stocks & 1 week & 1s & Exp & \begin{tabular}{@{}r@{}}Fixed decay\\Random Graph\end{tabular}\\ 
\cite{mastromatteo2011criticality} & Mastromatteo {\em et al.} & 100 & Trades & Stocks & 1 year & 1s & Exp & \\ 
\cite{Rambaldi:2014aa} & Rambaldi {\em et al.} & 1 & Quotation & FX & 1 year & 0.1s & N-Exp & News\\ 
\cite{Sahalia:2010hx} & Sahalia {\em et al.} & 1-2 & Price & Indices & $[22,33]$ years & 1day & Exp & Contagion\\ 
\end{longtable}